\def\expec#1{\langle#1\rangle}
\def\bigexpec#1{\left\langle#1\right\rangle}

\def\etal{{\frenchspacing\it et al.}}
\def\ie{{\frenchspacing\it i.e.}}
\def\eg{{\frenchspacing\it e.g.}}

\def\rms{{\frenchspacing r.m.s.}}

\def\pp{\noindent\parshape 2 0truecm 8.8truecm 0.3truecm 8.5truecm}  
\def\rf#1;#2;#3;#4 {\par\pp#1, #2, #3, #4 \par}
\def\rg#1;#2;#3;#4;#5 {\par\pp#1, #2, #3, #4 (#5)\par}
\def\rn{\pp}

\def\beq#1{\begin{equation}\label{#1}}
\def\eeq{\end{equation}}
\def\beqa#1{\begin{eqnarray}\label{#1}}
\def\eeqa{\end{eqnarray}}
\def\eq#1{equation~(\ref{#1})}
\def\Eq#1{Equation~(\ref{#1})}
\def\eqn#1{~(\ref{#1})}

\def\bfig{\begin{figure}[h] \centerline{\hbox{}}\vfill}
\def\efig{\end{figure}\vfill\newpage}

\def\sec#1{Section~\ref{#1}}
\def\Sec#1{Section~\ref{#1}}

\def\spose#1{\hbox to 0pt{#1\hss}}
\def\simlt{\mathrel{\spose{\lower 3pt\hbox{$\mathchar"218$}}
     \raise 2.0pt\hbox{$\mathchar"13C$}}}
\def\simgt{\mathrel{\spose{\lower 3pt\hbox{$\mathchar"218$}}
     \raise 2.0pt\hbox{$\mathchar"13E$}}}
\def\simpropto{\mathrel{\spose{\lower 
3pt\hbox{$\mathchar"218$}}
     \raise 2.0pt\hbox{$\propto$}}}

\def\tr{\hbox{tr}\>}
\def\realpart{\hbox{Re}\,}
\def\Mpc{{\rm Mpc}}

\def\nbar{{\bar n}}
\def\Nbar{{\bar N}}

\def\i{i}	
\def\ith{i^{th}}
\def\a{\alpha}	
\def\b{\beta}	
\def\bt{\tilde{b}}
\def\dirac{\delta^D}
\def\vk{{\bf k}}
\def\r{{\bf r}}
\def\vb{{\bf b}}
\def\k{{\bf k}}
\def\p{{\bf p}}
\def\pest{\tilde{\bf p}}
\def\q{{\bf q}}
\def\r{{\bf r}}
\def\x{{\bf x}}
\def\vx{{\bf x}}
\def\y{{\bf y}}
\def\z{{\bf z}}
\def\A{{\bf A}}
\def\B{{\bf B}}
\def\C{{\bf C}}
\def\E{{\bf E}}
\def\F{{\bf F}}
\def\I{{\bf I}}
\def\LL{{\bf\Lambda}}
\def\vtheta{{\bf\Theta}}

\def\vmu{{\bf m}}
\def\N{{\bf N}}

\def\PP{{\bf\Pi}}
\def\Veff{V^{eff}}

\def\Pt{{\tilde P}}
\def\S{{\bf S}}
\def\U{{\bf U}} 
\def\Z{\U} 
\def\Zt{\tilde{\Z}}
\def\u{{\bf u}}

\def\vzero{{\bf 0}}
\def\phih{\widehat{\phi}}
\def\psih{\widehat{\psi}}
\def\zh{\widehat{z}}
\def\nnorm{a}
\def\nnormh{\widehat{a}}
\def\vnnorm{{\bf a}}

\def\M{{\bf M}}

\def\Pt{{\tilde P}}

\def\delt{\delta_r}
\def\delth{\widehat{\delta}_r}
\def\deltash{\widehat{\delta}_s}
\def\vkh{\widehat{\vk}}
\def\rh{\widehat{\bf r}}
\def\Eh{\widehat{E}}
\def\tr{\hbox{tr}\>}
\def\dV{{d^3k\over (2\pi)^3}}
\def\dag{^\dagger}

\def\ed{\end{document}}

\def\RatingTable{
\begin{tabular}{lccc}
\hline
     &TRAD		&LIN	&QUAD\\
\hline
{\footnotesize Optimal on largest scales}	&$-$	&$+$	&$+$\\
{\footnotesize Optimal on smallest scales}	&$+$	&$-$	&$-$\\
{\footnotesize Simple \& uncorrelated errors}	&$-/+$	&$+/-$	&$+$\\
{\footnotesize Measures $z$-space distortions}	&$-$	&$+$	&$-$\\
{\footnotesize Accomodates systematic effects}	&$-$	&$+$	&$+$\\
\hline
\end{tabular}
\smallskip

\noindent
{\bf Table 1:} {\it Pros and cons of the traditional, 
linear (KL) and quadratic power spectrum estimation methods.}
\label{RatingTable}
}

\documentstyle[emulateapj,epsf,rotate]{article}
\begin{document}


%



\journalid{337}{15 January 1989}
\articleid{11}{14}


\submitted{To appear in {\it ApJ}, {\bf 499} (1998)}

\title{MEASURING THE GALAXY POWER SPECTRUM WITH FUTURE REDSHIFT SURVEYS}

\author{Max Tegmark\footnote{Hubble Fellow.}}

\affil{Institute for Advanced Study, Princeton, NJ 08540; 
max@ias.edu}


\author{Andrew J. S. Hamilton}

\affil{JILA and Dept. of Astrophysical, Planetary and 
Atmospheric Sciences,}

\affil{Box 440, Univ. of Colorado, Boulder, CO 80309, USA; 
ajsh@dark.colorado.edu}



\author{Michael A. Strauss\footnote{Alfred P. Sloan Foundation
Fellow.}$^,$\footnote{Cottrell Scholar of Research Corporation.}}

\affil{Department of Astrophysical Sciences, Princeton 
University, Princeton, NJ 08544; strauss@astro.princeton.edu}

\author{Michael S. Vogeley$^1$}

\affil{Department of Astrophysical Sciences, Princeton 
University, Princeton, NJ 08544; vogeley@astro.princeton.edu}

\and

\author{Alexander S. Szalay}

\affil{Department of Physics and Astronomy, Johns Hopkins 
University, Baltimore, MD 21218; szalay@pha.jhu.edu}



\begin{abstract}
Precision measurements of the galaxy power spectrum $P(k)$ 
require a data analysis pipeline that is both fast enough to be
computationally feasible and accurate enough to take 
full advantage of high-quality data.  
We present a rigorous discussion of different methods of power
spectrum estimation, with emphasis on the traditional Fourier method,
and linear (Karhunen-Lo\`eve; KL), and quadratic data compression
schemes, showing in what approximations they give the same 
result. 
To improve speed, we show how many of the
advantages of KL data compression and power spectrum estimation may be
achieved 
with a computationally faster quadratic method.
To improve accuracy, we derive analytic expressions for handling the 
integral constraint, since it is crucial that finite volume effects are 
accurately corrected for on scales comparable to the depth of the
survey.
We also show that for the 
KL and quadratic techniques, 
multiple constraints can be included via simple matrix operations,
thereby rendering the results less sensitive to galactic
extinction and mis-estimates of the radial selection function.
We present a data analysis pipeline that we argue 
does justice to the increases in both quality and quantity 
of data that upcoming redshift surveys will provide.
It uses three analysis techniques 
in conjunction: a traditional Fourier approach 
on small scales, a pixelized quadratic matrix method on large
scales and a pixelized KL eigenmode
analysis to probe anisotropic effects such as
redshift-space distortions. 
\end{abstract}


%
%



\makeatletter
\global\@specialpagefalse
\def\@oddfoot{
\ifnum\c@page>1
  \reset@font\rm\hfill \thepage\hfill
\fi
\ifnum\c@page=1
{\sl 
Available in color from
h t t p://www.sns.ias.edu/$\tilde{~}$max/galpow.html}
\hfill\\
\fi
} \let\@evenfoot\@oddfoot
\makeatother



\section{INTRODUCTION}

Observational data on galaxy clustering are rapidly 
increasing in both 
quantity and quality, which brings new challenges to data analysis.
As for quantity, 
redshifts had been published for a few 
thousand galaxies
15 years ago. Today the number is $\sim 10^5$ (Huchra, private communication), 
and ongoing 
projects such as
the AAT 2dF Survey and the Sloan Digital Sky Survey 
(hereafter SDSS;
see Gunn \& Weinberg 1995)
will raise it to $10^6$ in a few years.
Comprehensive reviews of past redshift surveys
are given by Efstathiou (1994), 
Vogeley (1995), 
Strauss \& Willick (1995) and Strauss (1997), the last also 
including a
detailed description of 2dF and SDSS. 
As for quality, 
more accurate and uniform photometric 
selection criteria
(enabled by {\eg} the well-calibrated 5-band photometry of 
the SDSS)
reduce potential systematic errors.
This increased data quality makes it desirable to avoid 
approximations in the data analysis 
process and to use methods that can constrain cosmological 
quantities as accurately as possible.  This is 
especially important since a wide variety of models currently appear to be
at least marginally consistent with the present data
(Peacock 1997; White {\etal} 1996; Vogeley 1997), so smaller error
bars on the power spectrum will be needed to discriminate between them.
However, the increased
data quantity makes it a real challenge to perform such
an accurate analysis;
as we will discuss at some length, 
a straightforward application of any such
method is computationally unfeasible for data sets
as large as those from 2dF and SDSS.

This paper addresses both the accuracy and 
speed issues for measuring the galaxy power spectrum $P(k)$. 
For the highest accuracy, we advocate the use of lossless pixelized
data compression 
methods, both linear (Karhunen-Lo\`eve) and quadratic.  
To improve the speed, we 
present a fast implementation of a pixelized quadratic power  
estimation
method, and show how it can reproduce Karhunen-Lo\`eve
results (Vogeley \& Szalay 1996; 
Tegmark, Taylor \& Heavens 1997 --- hereafter TTH)
exactly, without the need to solve eigenvalue problems,
although it does not have the ability to measure redshift-space
distortions. 

The rest of the paper is organized as follows.
Section~\ref{WishListSec} is a ``buyers guide", where we
list the various properties that are important 
when deciding which data analysis method to use.
In \sec{MethodsSec}, we review the existing methods 
for power spectrum estimation in a common framework, make various extensions
of them and present the fast quadratic data compression and power
spectrum technique. 
\Sec{RelationSec} describes how the various methods are interrelated,
and
\sec{SystematicsSec} discusses how they can be immunized from
various systematic effects, such as errors in the extinction model;
many of the technical details are given in Appendices A and B. 
\Sec{ProsConsSec} discusses the pros and cons of the different techniques.
We conclude that to meet all criteria on the
wish list of \sec{WishListSec}, it is necessary 
to combine three of the principal methods described. 
The data analysis pipeline that we propose is summarized in 
Figure~1, and the reader may wish to glance at this 
before delving into the 
details of \sec{MethodsSec}, to obtain an overview of how 
everything fits together.  We ignore redshift-space distortions for
most of this paper, and assume that clustering is isotropic.  In
Section~\ref{SystematicsSec} and Appendix C, we discuss this issue
further. 

\bigskip

\section{COMPARING METHODS: A WISH LIST}

\label{WishListSec}

Since the ultimate goal of the analysis of clustering
in a galaxy redshift survey is to constrain cosmological 
models, 
we want a method that minimizes the statistical error 
bars on cosmological parameters\footnote{
This is not a mere unimportant detail, since 
doubling the error bars (which an inferior
method can easily do) is comparable to reducing the survey volume and the
number of galaxies probed by a factor of four.
}
and is robust against potential systematic errors.
We will now discuss the former issue in some detail, and 
return to the latter in \sec{SystematicsSec} and Appendix B. 

\subsection{The problem to be solved}

\label{ProblemSec}

The data set from a galaxy redshift survey consists of 
$N$ three-dimensional vectors $\r_\a$ ($\a=1,...,N$) 
giving the measured positions 
of 
the galaxies.  
Following Peebles (1973, 1980), we model these positions as generated
by a random Poissonian point process where the galaxy density is modulated
by both selection effects and fluctuations in the underlying
matter distribution. The former are described
by $\nbar(\r)$, the {\it selection function} 
of the galaxy
survey under consideration, defined as the expected
galaxy density. Thus $\nbar(\r)dV$
is the expected (not the observed) number of galaxies in 
a volume $dV$ about $\r$ in the absence of clustering.
This function typically falls off at large distances,
and can exhibit small angular variations due to extinction.
It vanishes outside the survey volume.
The fluctuations in the underlying matter density are
given by the field $\delt(\r)$, which is 
not to be confused with the
Dirac delta function $\dirac$.
This means that the observed galaxy distribution 
$n(\r) = \sum_\alpha \dirac(\r-\r_\a)$
is modeled as a 3D Poisson process
with intensity $\lambda(\r) = \nbar(\r)[1+\delt(\r)]$.

The density fluctuations $\delt$ are modeled as 
a homogeneous and isotropic 
(but not necessarily Gaussian) 
random field. This implies that the Fourier transform $\delth$
of the density fluctuation field obeys the simple relation
\beq{PdefEq}
\expec{\delth(\k)^*\delth(\k')} = (2\pi)^3 \dirac(\k-\k')P(k)
\eeq
for some function $P$ which depends only on the magnitude of $\k$,
not on its direction\footnote{Note that \eq{PdefEq} only holds if
positions are measured in real space.  Redshift-space distortions
couple modes of different \k; see Appendix C for more details.} 
$P$ is known as the {\it power spectrum}. 
Because of \eq{PdefEq}, $P$ contains all the information 
needed to compute any statistical quantities that depend quadratically
on $\delt$, for instance the variance or the correlation function on different
scales. Moreover, if the random field $\delt$ is {\it Gaussian} (which 
means that the joint probability distribution of $\delt$ at any number of
points is a multivariate Gaussian), then $P$ characterizes 
$\delt$ {\it completely} and contains all the information
needed to answer any statistical question whatsoever about $\delt$.
Inflationary theory (e.g., Peebles 1993) and observations of
large-scale structure (e.g., Strauss \& Willick 1995) imply that
$\delt$ is Gaussian 
on large scales, making $P$ an important quantity in cosmology.
The power spectrum estimation problem, which is the topic of this paper, 
is to estimate $P(k)$ given the observed realization of $n(\r)$.

\subsection{The traditional approach}

\label{TradApproachSec}

Let us parameterize the power spectrum $P(k)$ by some set of 
parameters
$\theta_i, i = 1, 2, ...$, grouped into a vector $\vtheta$. 
These may be either
the band powers in a set of narrow bands, or physically 
motivated parameters such
as the normalization $\sigma_8$, the shape parameter 
$\Gamma$, the 
primordial spectral index $n$, {\it etc}.
Let us package our data set into a vector $\x$; much of the
distinction between different methods discussed in \sec{MethodsSec}
lies in the way this packaging is done.
The standard approach to parameter estimation is to write down the
expression for the probability distribution $f(\x;\vtheta)$.
Here we interpret $f$ as a probability distribution over 
$\x$ for a fixed $\vtheta$.
In a Bayesian statistical analysis with a uniform prior 
probability distribution for $\vtheta$, one
reinterprets $f$ as a probability distribution 
over $\vtheta$ for a given data set $\x$, and to clarify this 
distinction renames $f$ the likelihood function.
The final results are often presented as 
contour plots of this likelihood function, 
as at the bottom of Figure~1.

If we take $\x$ to be the raw data set, {\ie}, the measured coordinates
$\r_\a$ ($\a=1,...,N$)
of the $N$ measured galaxies, then the likelihood function 
$f$ is unfortunately hopeless to compute numerically, since it involves the 
$N$-point correlation function.
Even in the Gaussian approximation that $f$ is given by a product over two-point
correlation functions (\eg, White 1979; Fry 1984), 
this 
requires evaluating a multivariate
polynomial of degree $N/2$ in the correlations of the $N(N+1)/2$ galaxy 
pairs, and the CPU time required for this grows faster than exponentially
with $N$ (Dodelson, Hui, \& Jaffe 1997).
The traditional approach has therefore been to take $\x$ to be 
something else: 
band-power estimates of the power spectrum.
These are essentially computed by multiplying the observed density field
by some weight function, Fourier transforming it, taking the squared modulus
of the result and averaging over shells in $k$-space (\sec{TraditionalSec}).
In the (sometimes poor) approximation that the probability distribution for 
$\x$ is a multivariate 
Gaussian, its probability distribution is 
\beq{GaussianLikelihoodEq}
f(\x;\vtheta) \propto 
|\C|^{-1/2}\,e^{-{1\over 2}(\x-\vmu)\dag\C^{-1}(\x-\vmu)},
\eeq
where $|\C|$ denotes the determinant of $\C$, and the mean vector 
$\vmu\equiv\expec{\x}$ and the covariance matrix 
$\C\equiv\expec{\x\x\dag}-\vmu\vmu\dag$ depend 
on $P(k)$ and hence on the unknown parameters $\vtheta$.
Much of the model testing to date has been rather 
approximate, often little more than a ``$\chi$-by-eye'' fit of
theoretical power spectra to the data, which is tantamount to ignoring the
correlations between the power estimates (the off-diagonal elements of $\C$).
This approach clearly does not utilize all the information present in the data,
and can also bias the results.

If we had infinite computer resources, we would improve the situation
by simply performing
an exact brute force likelihood analysis on the raw data set. 
Is there a faster way of obtaining the same result? 


\subsection{The notion of a lossless method}

We will call a method for analyzing a data set {\it 
unbeatable\/}
or {\it optimal\/} if no other method can place tighter 
constraints 
on cosmological models (as parameterized by $\vtheta$)
using this data.  In this paper, we are focussed on the power spectrum
measured in real space, and so we restrict ourselves to measurements
of the parameters which determine the power spectrum itself. 

\subsubsection{The Fisher Information Matrix}

This can be made 
precise using the formalism 
of the {\it Fisher information matrix}
(see TTH for a comprehensive review of this application),
which offers a simple and a useful way of measuring how much
information each step in the pipeline of Figure 1 destroys.
Given any set of cosmological parameters of interest
denoted $\theta_i$, $i=1,2,...$, 
their Fisher matrix $\F$ gives the smallest error bars 
with which the parameters
can possibly be measured from a given data set.
If the probability distribution for the data set 
given the parameter values is
$f(\x;\vtheta)$, then the Fisher matrix is defined by 
(Fisher 1935)
\beq{FisherDefEq}
\F_{ij} \equiv - \bigexpec{{\partial^2 \ln 
f\over\partial\theta_i\partial\theta_j}}.
\eeq
Crudely speaking, $\F^{-1}$ can be thought of as the
best possible covariance matrix for the measurement errors on 
the parameters.
Indeed, the Cram\'er-Rao inequality 
(Kenney \& Keeping 1951; Kendall \& Stuart 1969)
states
that no unbiased method whatsoever
can measure the $\ith$ parameter with 
error bars (standard deviation) 
less than $1/\sqrt{\F_{ii}}$.
If the other parameters are not known but are estimated
from the data as well, the minimum standard deviation
rises to $(\F^{-1})_{ii}^{1/2}$.
This formalism has recently been used to assess the accuracy 
with which cosmological parameters can be measured
from future galaxy surveys 
(Tegmark 1997b; Goldberg \& Strauss 1998; Hu, Eisenstein, \& Tegmark 1997)
and cosmic microwave background experiments 
(Jungman {\etal} 1996; Bond, Efstathiou, \& Tegmark 1997; Zaldarriaga,
Seljak, \& Spergel 1997).

\subsubsection{Checking for leaks in the pipeline}

By computing the Fisher matrix separately from each of the 
intermediate data
sets in Figure 1, we can track the flow of information
down the data pipeline and check for leaks.
For instance, if the Fisher matrix computed from the raw 
positional data
is identical to that computed from the (much smaller) data 
set consisting of
the band power estimates, then
the power spectrum estimation method is {\it lossless}
in the sense that no information about the parameters of interest
has been lost in the process of compressing the data set from, 
say, $10^6$ numbers down to $50$; cf., the discussion in \sec{CompressSec}. 
We will use this criterion when comparing
different data analysis techniques below. 

\subsubsection{The power spectrum Fisher matrix}

Whether a method is lossless or not generally depends on 
which 
parameters we are interested in estimating.
Fortunately, as shown by Tegmark (1997a, hereafter T97), 
certain methods can be shown to be 
lossless for {\it any} set of parameters in a large class.
An important special case are quantities that parameterize the
power spectrum $P(k)$, such as $\sigma_8$ and $\Gamma$; 
all the information on these parameters
is retained if the power spectrum itself (parameterized by 
the power in many narrow bands) can be measured with 
the minimal error bars. This means that one can test whether
a method is lossless simply by computing the Fisher matrix 
for the band powers.  This also means that band powers have a 
special status compared to other parameters: if we simply measure $P(k)$ 
as accurately as possible, this measured function will retain all the 
information about all cosmological parameters.
All this is of course only true within the framework of
Gaussian and isotropic 
models, which are uniquely characterized by their power spectrum.
Specifically, these methods {\it do\/} 
lose information
about parameters that affect the data set not only via $P(k)$.
Important examples to which we will return are redshift-space
distortions and uncorrected galactic extinction, 
both of which introduce differences between the angular
and radial clustering patterns. On small scales, nonlinear clustering
creates non-Gaussian fluctuations where in addition to $P(k)$,
higher order moments also contain cosmological information.

\subsection{Data compression, simplicity and speed}
\label{CompressSec}

A second 
and rather obvious 
criterion for comparing data 
analysis methods is their numerical feasibility. 
For instance, the
brute-force likelihood analysis of the raw data set (the galaxy positions)
described above is lossless, but
too time-consuming to be numerically feasible
when 
$N$, the number of galaxies, is large. 

When the brute-force method is unfeasible, the general 
approach is
to perform some form of {\it data compression}, whereby the 
data set
is reduced to a smaller and simpler one which is easier to 
analyze.
If the data compression step is lossless, a brute-force 
analysis
on the compressed data set clearly gives just as small error 
bars
as one on the raw data would have done.

To facilitate parameter estimation further down the 
pipeline, it is 
useful if the statistical properties of the power
spectrum estimates are simple and easy to calculate.
The simplest case (which often occurs with 
the pixelized methods described below) is that where 
the data set is a multivariate Gaussian, described by the
likelihood function of \eq{GaussianLikelihoodEq}.
Then the slowest step in the likelihood calculation is 
computing the
determinant of the $N\times N$ covariance matrix $\C$, for which 
the CPU time
scales as $N^3$, so it is desirable to make the compressed 
data 
set as small as is possible without losing 
information\footnote{Another problem with large $N$ is the memory 
requirements for an $N\times N$ covariance matrix; if $N = 10^5$, for
example, this is currently beyond the RAM capacity of all but the very
largest computers.}. 
It is also convenient if the statistical properties
of the compressed data set, in particular, its covariance matrix,
can be computed analytically. We will see that this is the
case for the Karhunen-Lo\`eve and quadratic data compression methods
describe below, but not for the  
the maximum-likelihood method, where it
requires numerical computation of the entire likelihood surface. 
Finally, the simplest covariance matrix one can desire
is clearly one that is diagonal, {\ie}, where the errors on 
the elements
of the compressed data set are uncorrelated. 

\subsection{The wish list}

\noindent
In summary, the ideal data analysis/data compression method would
\begin{enumerate}
\item be lossless, at least for the parameters of interest,

\item give easy-to-compute and uncorrelated errors,

\item be computationally feasible in practice,

\item allow one to account for redshift-space distortions and
systematic effects. 

\end{enumerate}
The first two items can be summarized by saying 
that we want the method
to retain the cosmological information content 
of the original data set, 
distilled into a set of mutually exclusive (2) 
and collectively exhaustive (1) chunks. 
In other words, the information chunks should be independent 
and together retain all the information from the original data set

\section{POWER SPECTRUM ESTIMATION METHODS --- AN OVERVIEW}
\label{MethodsSec}

In this section, we review the various methods 
for power spectrum estimation that have been proposed in the 
literature and present various extensions.  We start in
\sec{NotationSec} by developing a
formalism that is common to all our approaches. In
\sec{PixelizationSec}, we discuss how the data might be discretized;
that is, different ways of packaging the data into a convenient form $\x$.  
We then discuss various methods of power spectrum estimation:
traditional methods which take the square of the amplitude 
of the Fourier modes (\sec{TraditionalSec}), the method of brute force
likelihood (\sec{BruteSec}), the linear Karhunen-Lo\`eve data
compression method 
(\sec{KLsec}), and a new quadratic data compression method (\sec{QuadraticSec}).  Our
suggested approach to power spectrum estimation involves a combination
of these methods in different regimes, as summarized in Figure~1, and
described in more detail in \sec{DiscussionSec}. Throughout \sec{MethodsSec},
we ignore redshift-space distortions and other systematic effects, and
assume that clustering is isotropic.  We return to this topic in
\sec{SystematicsSec}.

\subsection{Density field, shot noise and window functions} 
\label{NotationSec}

With the exception of the brute force maximum likelihood technique, all of
the methods described below compute band power  
estimates $q_\i$ that are quadratic functions of the density 
field $n$, which means that they can be written as
\beq{DirectQuadEq}
q_\i = \int\int E_\i(\r,\r')
{n(\r)\over\nbar(\r)}
{n(\r')\over\nbar(\r')} d^3r\,d^3r' 
= 
\sum_{\a,\b} {E_\i(\r_\a,\r_\b)\over\nbar(\r_\a)\nbar(\r_\b)}
\eeq
for some real-valued symmetric pair weighting functions $E_\i$ which are 
designed
to isolate different ranges of wavenumber $k$ --- the 
methods simply differ in their choices of $E_\i$.
Taking the expectation value of $n(\r)n(\r')/\nbar(\r)\nbar(\r')$ produces three terms:
``1" from the mean density, $\dirac(\r-\r')/\nbar(\r)$ from shot noise
and $\delt(\r)\delt(\r')$ from density fluctuations.
By a derivation analogous to FKP and T95, one finds that these three terms give
\beq{GenQuadExpecEq}
\expec{q_\i} = W_\i(\vzero) + b_\i + \int
W_\i(\k)P(k)\dV,
\eeq
where $W_\i$, the three-dimensional {\it window functions}, are given by
\beq{3DwindowEq} 
W_\i(\k) = \Eh_\i(\k,\k)
\eeq
and 
\beq{EhDefEq}
\Eh_\i(\k,\k')\equiv
\int E_\i(\r,\r')
e^{-i\k\cdot\r}
e^{i\k'\cdot\r'}d^3r\>d^3r'
\eeq 
is a Fourier 
transform of $E_\i$. We will often find it convenient to rewrite the
last term of equation~(\ref{GenQuadExpecEq}) as $\int_0^\infty
W_\i(k)P(k)dk$, where the {\it one-dimensional\/} window function is
the angular average
\beq{GenQuadWinEq}
W_\i(k) = k^2\int W_\i(\k) d\Omega_k.
\eeq
The {\it shot noise bias} is given by
\beq{GenQuadBiasEq}
b_\i\equiv\int{E_\i(\r,\r)\over\nbar(\r)}d^3r.
\eeq
Alternatively, $b_\i$ can be made to vanish by omitting the 
terms with $\a=\b$ from the double sum in \eq{DirectQuadEq},
as described in Appendix A. 
The term $W_\i(\vzero)$ simply probes the mean density of
the survey, and as described in \Sec{SystematicsSec} and Appendix
B, the functions 
$E_\i$ should always be chosen such that this term vanishes,
{\ie}, so that $\int E_\i(\r,\r') d^3r\>d^3r'=0$, thereby
immunizing the power estimates to 
errors in normalization of $\nbar$ 
(\Sec{ConstraintDiscussionSec})\footnote{
In fact, the stronger constraint $\int E_\i(\r,\r') d^3r'=0$ for all $\r$
should be enforced, as discussed in Appendix B.
}. 
The desirability of choosing windows with this property was first
explicitly pointed out by Fisher \etal\ (1993), and this prescription
was used also by {\eg} Hamilton (1992)
and Cole, Fisher \& Weinberg (1994).
We want to interpret $q_i$ in \eq{GenQuadExpecEq} as probing 
a weighted average of the power spectrum, 
with the window function giving the weights, so
$E_\i$ should be normalized so that $W_\i(\k)$ 
integrates to unity
(throughout this paper, we will write the volume element in
Fourier space as $d^3k/(2\pi)^3$ rather than $d^3k$,
since this minimizes the number of $2\pi$-factors
elsewhere). Using \eq{3DwindowEq} and
Parseval's theorem, this gives
\beq{EnormEq}
1=\int_0^\infty W_i(k){dk\over (2\pi)^3}=\int W_i(\k)\dV=\int E_\i(\r,\r)d^3r,
\eeq
so if we think of $E_\i(\r,\r')$ as a matrix with indices
$\r$ and $\r'$, 
this normalization condition is simply $\tr[E_\i]=1$. 

As described in Hamilton (1997a), the window function has a simple 
geometrical interpretation. Let us rewrite \eq{GenQuadWinEq} as
\beq{GenQuadWinEq2}
W_\i(k) = {4\pi k^2}\int_0^\infty w_\i(r)j_0(kr)dr,
\eeq
where the {\it separation weighting} is defined by  
\beq{PairSepDefEq}
w_\i(d)\equiv\int\!\int E_\i(\r,\r')\dirac(|\r-\r'|-d)d^3r\>d^3r'.
\eeq
The separation weighting is the average of $E_i$ 
over all pairs of points separated by a fixed distance $d$,
weighted by the number of such pairs.
\Eq{GenQuadWinEq2} shows that the  
only aspect of $E_\i$ that affects the window
function is the separation weighting, since two different $E_i$ that
give the same $w_\i$ will produce identical window functions.

An important special case, to which we return in the next subsection,
is that where $E_\i$ is of rank one, {\ie}, of the separable form
\beq{RankOneEq}
E_\i(\r,\r') = \psi_\i(\r)\psi_\i(\r')^*
\eeq
for some $\psi_\i$. 
In this case, 
\eq{DirectQuadEq} can be written as $q_i=|x_i|^2$,
where 
$x_i=\int\psi_\i(\r) n(\r)/\nbar(\r)$ (see \eq{xDefEq} below),
and the window function becomes simply 
\beq{RankOneWinEq}
W_\i(\k) =|\psih_i(\k)|^2.
\eeq
(Here and throughout, hats denote Fourier transforms;
$\psih(\k)\equiv\int e^{-i\k\cdot\r}\psi(\r)d^3r$ --- see Appendix D.)
We will see that both
the traditional methods (\sec{TraditionalSec})
and the KL method (\sec{KLsec})
are of this separable form,
while the quadratic data compression method that we 
present in \sec{QuadSec} is not.
We will occasionally refer to the former methods as {\it linear} 
data compression, since they begin by taking linear combinations
of the data with weight functions $\psi_\i(\r)$; cf.,
\eq{xDefEq}.
The latter method, on the other hand, is intrinsically quadratic since
$q_i$ is not merely the square of some quantity that is linear in the data.
This is because
its pair weighting $E$ is optimized to provide a minimum 
variance power estimator,
which makes $E$ a quadratic form of rank greater than one. 

To avoid confusion, the reader should bear in 
mind that when we distinguish between linear and quadratic methods below, 
we are referring to linear versus quadratic {\it data compression}.
The power estimator $q_i$ is of course quadratic in both cases.

\subsection{Pixelization}

\label{PixelizationSec}

Hamilton (1997a) has recently derived the functions 
$E_\i$ that provide the minimum-variance power 
spectrum estimates for an arbitrary selection function and
survey geometry. Unfortunately, this optimal weighting
scheme is in general impractical to implement numerically,
as it involves a computationally cumbersome infinite series 
expansion --- only in the small-scale limit does it become 
simple, as will be described in \sec{SmallScaleSec}. 
To proceed numerically, it is therefore convenient to 
discretize the problem. 
This reduces it to one similar to that
of cosmic microwave background (CMB) experiments:
estimating a power spectrum given noisy fluctuation 
measurements
in a number of discrete ``pixels". Once the pixelization is done,
the remaining  
steps are 
quite analogous to the CMB case (T97), and involve mere 
matrix operations such as inversion and diagonalization. 
These operations can often be further simplified by a more 
suitable choice of pixelization.

Let us define the overdensity in $N$ ``pixels" $x_1,...,x_N$ 
by 
\beq{xDefEq}
x_i \equiv\int\left[{n(\r)\over\nbar(\r)}-1\right]\psi_i(\r) d^3r
\eeq
for some set of functions $\psi_i$.
We discuss
specific choices of 
$\psi_i$ in some detail below. One generally strives 
either to make these functions fairly localized in real space
(in which case the pixelization is a generalized form of 
counts in cells) or fairly localized in Fourier space 
(in which case we will
refer to the functions $\psi_i$ as ``modes" and to $x_i$ as 
expansion coefficients). 
The reader should be warned that the latter case is far from 
what would normally be thought of as ``pixels'', and that we
will be using this terminology nonetheless to stress 
that all discretization schemes can be treated in a 
mathematically equivalent way.

The ``$-1$" term in \eq{xDefEq}
simply subtracts off the mean density.  As we discuss in
\sec{SystematicsSec} and Appendix B, one can and should choose the weight
functions $\psi_i$ to have zero mean, making this term irrelevant
(note indeed our use of this fact immediately preceding
\eq{RankOneWinEq}). 
This corresponds to requiring that
\beq{psiMeanEq} 
\int \psi_i(\r) d^3r = 0. 
\eeq

Let us group the pixels $x_i$ into an $N$-dimensional vector 
$\vx$. From 
equations\eqn{xDefEq} and\eqn{psiMeanEq} and a
generalization of the derivation of \eq{GenQuadExpecEq},
T95 shows that
\beqa{xExpecEq}
\expec{\vx}&=&\vzero,\\
\label{xCovEq}
\expec{\vx\vx^\dagger}&=&\C\equiv\N+\S,
\eeqa
where the shot noise covariance matrix is given by
\beq{NdefEq}
\N_{ij} = \int {\psi_i(\r)\psi_j(\r)^*\over\nbar(\r)}d^3 r
\eeq
and the signal covariance matrix is 
\beq{SdefEq}
\S_{ij} = \int \psih_i(\vk)\psih_j(\vk)^*
P(k)\dV.
\eeq


How should we choose our $\psi_i$'s to pixelize space? 
For a pixelization to be useful,
we clearly want the data set $\vx$ to retain as large a 
fraction as possible 
of the cosmological information from the original data set 
(the 
galaxy positions), while simultaneously simplifying 
subsequent calculations. We here list several natural options, 
most of which have appeared in the literature. 


\subsubsection{Counts in cells}

Here one partitions the survey volume into
$N$ mutually exclusive and collectively exhaustive volumes 
$V_i$, and defines $\psi_i(\r)=\nbar(\r)$ if $\r\in V_i$, 
$\psi_i(\r)=0$ elsewhere. Thus $x_i$ is simply the number of 
galaxies observed in $V_i$, minus the expected average. 
A set of useful approximations for computing $\N$ and $\S$
for this case is derived in VS96. To 
keep the number of cells from becoming intractably large, 
one might choose the cells to be larger in distant and poorly 
sampled regions of space than nearby.


With these sharp-edged cells, any linear combinations of 
pixels will correspond to a weight function that is
discontinuous at cell boundaries. To 
avoid power leakage problems
that this can in principle cause, one might use cells 
with ``fuzzy" boundaries instead, for instance Gaussians as 
described in T95.


One can greatly simplify the computation of the covariance matrix 
by choosing all cells to have the same shape and to be 
spherically symmetric (\eg, spheres or Gaussians),
since 
the resulting $\S_{ij}$ will depend only on the separation of the pixel 
centers and this correlation function can be pre-computed and splined 
once and for all. 
For the volume-limited case ($\nbar$ constant), performing 
the appropriate
integrals for identical 
spherical cells of radius $R$, separated by $uR$, gives
\beq{SphereNeq}
\N_{ij} = \left \{\begin{array}{ll}
    {(2-u)^2(4+u)\over 16}\nbar V &{\rm if}\ u < 2, \\
    0                             &{\rm if}\ u \ge 2,
	           \end{array}
\right.
\eeq
where $V=4\pi R^3/3.$

\subsubsection{Fourier modes}

All of the above-mentioned pixels were fairly well-localized 
in real space. To 
make pixels reasonably localized in Fourier space, 
one can choose modes $\psi_i$ that are plane waves, tapered 
by some weight function $\phi$ to make them square 
integrable:
\beq{FourierPixelizationEq}
\psi_i(\r) = \phi(\r)e^{i\k_i\cdot\r}
\eeq
for some grid points $\k_i$ in Fourier space. Choosing all 
modes that are periodic in a box containing the survey volume 
will ensure that this is a complete set, although some 
high-frequency cutoff is of course necessary to keep the number
of pixels finite.
Four different choices of the volume
weighting function $\phi$ have appeared 
in the literature:
\beqa{psiChoiceEq}
\phi(\r)&\propto&\left\{{\hbox{1 inside survey 
volume}\>\>\atop\hbox{0 outside survey volume}}\right.\\
\label{APMeq}
\phi(\r)&\propto&\nbar(\r)\\
\label{FKPeq}
\phi(\r)&\propto&{\nbar(\r)\over 1+\nbar(\r)P}\\
\label{T95eq}
\phi(\r)&\propto&\hbox{eigenfunction of }\left[\nabla^2-
{\gamma\over\nbar(\r)}\right]
\eeqa
All are to be normalized so that the corresponding 
window functions integrate to unity -- $\int\phi(\r)d^3r=1$ 
as shown below.
Note that without careful choice of the $\k_i$'s,
\eq{psiMeanEq} will {\it not\/} be satisfied in general 
for these pixelizations.
The first choice, which weights all volume elements in the 
survey equally, was employed by {\eg} 
Vogeley {\etal} (1992), Fisher {\etal} (1993), and
Park {\etal} (1994).
The second choice is used when determining the angular power spectrum
of a sample without redshifts, {\eg}, the APM survey
(Baugh \& Efstathiou 1994). Then  
all galaxies by default receive equal weight
(moreover, modes can of course only be computed in the 
directions perpendicular to the line of sight).
The third choice, advocated by Feldman, Kaiser \& 
Peacock (1994, hereafter FKP), minimizes the variance 
in the limit
when $k^{-1}\ll$ the depth of the survey
and is discussed in detail in \sec{FKPsec}.
Here $P$ denotes an {\it a priori} 
guess of the power
in the band under consideration.
The fourth choice (Tegmark 1995, hereafter T95), 
gives the narrowest window function for a given variance, where
the constant $\gamma$ determines the tradeoff.

\subsubsection{Spherical harmonic modes}

The spherical 
wave choice
\beq{HarmonicPixelEq}
\psi_i(\r) = Y_{\ell m}(\rh) j_\ell(k_n r),
\eeq
where $Y_{\ell m}$ is a spherical harmonic and $j_\ell$ a 
spherical Bessel function, is well-suited for full-sky redshift
surveys, and has the advantage (Fisher, Scharf, \& Lahav 
1994; 
Heavens \& Taylor 1995)
of greatly simplifying inclusion of the effect of 
redshift-space distortions in the analysis.

\subsubsection{Guessed eigenmodes}
\label{GuessedmodesSec}

In \sec{SNeigenmodeSec}, we describe a set of smooth functions 
known as continuum signal-to-noise eigenmodes. These modes 
have a number of useful properties.  In particular, 
there is an integer $N$ such that the first $N$ eigenmodes retain 
virtually all the cosmological information in the survey.
If one has a reasonable guess as to the shape of 
these functions {\it a priori}, before computing them exactly
(they depend on the survey geometry), 
the first $N$ of these guessed modes are obviously a good choice
for the pixelization functions $\psi_i$, since the
amount of information destroyed by the pixelization process will then be
small.

\subsection{Traditional methods}

\label{TraditionalSec}

The traditional approach has been to estimate the power 
by simply squaring the pixels ($q_i=|x_i|^2$), choosing the 
pixels to be Fourier modes as described above. This corresponds
to the pair weighting of \eq{RankOneEq}, where $\psi_i$ is given by
\eq{FourierPixelizationEq} and $\phi$ is specified by \eq{psiChoiceEq},
\eqn{APMeq}, \eqn{FKPeq} or~\eqn{T95eq}.
\Eq{GenQuadExpecEq} shows that 
\beq{TradqExpecEq}
\expec{q_i}=
\phih(\vzero)^2 + \int{\phi(\r)^2\over\nbar(\r)}d^3r + 
 {1\over(2\pi)^3}\int|\phih(\k-\k_i)|^2 P(\k)\dV, 
\eeq
the last term of which is 
simply the true power spectrum convolved 
with the function $|\phih(\k)|^2$.
The 3D window  function is thus 
$W_i(\k)=|\phih(\k-\k_i)|^2$. 
Using this and Parseval's theorem (or \eq{EnormEq}
directly), one finds that the window function normalization 
constraint $\int W_i(\k)d^3k/(2\pi)^3=1$ corresponds to simply
\beq{RankOneNormEq}
\int\phi(\r)^2 d^3r = 1.
\eeq
These simple expressions have frequently been used in the literature, 
but an annoying complication has often been neglected.    
As described in \sec{SystematicsSec}, the 
fact that the normalization of $\nbar$ 
is not known {\it a priori}, but 
is determined from the observed galaxies
(the so-called integral constraint problem) 
can be eliminated by choosing weight functions $\psi_i$ that are orthogonal
to the monopole, {\ie}, such that $\psih_i(\vzero)=0$. 
Although this is easy to arrange with the pixelized methods described
below, the choice of \eq{FourierPixelizationEq} does generally {\it
not} have this 
important property. To obtain a correct answer with the traditional
methods, this 
must be corrected for. As shown in Appendix B, the power estimator
\beq{PtDefEq}
\Pt_i \equiv 
\left[{|x_i|^2-b_i\over A_i}\right]
\eeq
is unbiased and incorporates the integral constraint correction
(when $\nbar$ is normalized so that $x_i=0$ for $\k_i=0$) if 
the normalization factor $A_i$ 
and 
the shot noise correction $b_i$ 
are given by 
\beqa{NormDefEq}
A_i&=&\left(1+\left|{\phih(\k_i)\over\phih(\vzero)}\right|^2
\right)a(\vzero)
- 2\,\realpart\left\{{\phih(\k_i)\over\phih(\vzero)} 
a(\k_i)\right\},\\
\label{sshotDefEq}
b_i&=&\left(1+\left|{\phih(\k_i)\over\phih(\vzero)}
\right|^2\right)b(\vzero)
- 2\,\realpart\left\{{\phih(\k_i)\over\phih(\vzero)} 
b(\k_i)\right\},
\eeqa
where the functions $a$ and $b$ are defined by
\beqa{abDefEq}
a(\vk)&\equiv&\int \phi(\r)^2 e^{i\vk\cdot\r}d^3r,\\
b(\k)&\equiv&\int{\phi(\r)^2\over\nbar(\r)}e^{i\vk\cdot\r}d^3r.
\eeqa
If the survey is volume limited, then $\nbar$ is independent 
of $\r$,
$b(\k)=a(\k)/\nbar$, and 
$b_i = A_i/\nbar$. 

After computing power estimates $q_i$ at a large grid of points 
$\k_i$, one finally takes some weighted averages of the $q_i$ to obtain
power estimates in some bands of (scalar) $k$ ($=|\k|$). 
The problem
of finding the weights  
that minimize the variance of the band power estimators is unfortunately quite
a difficult one, and in general involves solving a numerically unpleasant 
quadratic programming problem (T95). For this reason, the 
customary approach has been to simply give all $q_i$ equal weights in 
some spherical shells in $\k$-space although, as described in VS96, 
this is in general far from optimal.

\subsection{Brute force likelihood method}

\label{BruteSec}


Let us parameterize the power spectrum $P(k)$ by some 
parameter vector $\vtheta$ as in \sec{TradApproachSec}.
In the approximation that the probability distribution for the pixel
vector $\x$ is a multivariate Gaussian, it is given by \eq{GaussianLikelihoodEq}
with mean $\vmu=\vzero$.
The {\it maximum likelihood estimator} of $\vtheta$, denoted 
$\vtheta_{ml}$, is
simply that $\vtheta$-vector that maximizes the likelihood
$f(\x;\vtheta)$. 
This maximization problem can unfortunately not be solved 
analytically when the number of
pixels exceeds one, so $\vtheta_{ml}$ is a complicated non-linear
function of $\x$ that must 
be computed by solving the maximization problem numerically. 
Since one generally wants error bars on  
the estimate as well, one typically evaluates the likelihood 
function at a dense grid of
points in parameter space and rescales it to integrate to 
unity.
These final results are often illustrated in contour plots, 
as at the bottom of
Figure~1.

\subsection{Linear (Karhunen-Lo\`eve) method}

\label{KLsec}

As will be discussed in \sec{ProsConsSec}, 
the traditional methods generally destroy information, while the
brute force method is lossless but 
computationally impractical.
The Karhunen-Lo\`eve (KL) method (Karhunen 1947) 
maintains the advantage of the brute-force method 
(indeed, it can produce an essentially identical answer faster), 
and has additional useful features as well, as we detail below.
It was first introduced into large-scale structure analysis
by Vogeley \& Szalay (see VS96), and it has also been 
successfully
applied to Cosmic Microwave Background data ({\eg} Bunn 1995; 
Bond 1995; TTH; Bunn \& White 1997; Jaffe, Knox, \& Bond 1997).
Like any method designed to minimize error bars, the KL-technique
requires an {\it a priori} assumption for the power spectrum. 
This is referred to as the {\it fiducial} $P(k)$. As will be described 
in sections \ref{TroubleSpottingSec}, 
\ref{QuadMLrelSec} and 
\ref{RedHerringSec}, this is not a problem in practice, since
a bad fiducial model does not bias the result. Moreover,  
the measurement itself can be used as the fiducial model 
in an iterative procedure if desired.

We start by defining signal-to-noise eigenmodes in \Sec{SNeigenmodeSec},
before generalizing the technique in \Sec{GeneralKLsec}. 

\subsubsection{Signal-to-noise eigenmodes}
\label{SNeigenmodeSec}

The signal-to-noise eigenmode method consists of defining 
a new data vector
\beq{yDefEq}
\y\equiv\B\x,
\eeq
where $\vb$, the rows of the matrix $\B$, are the $N$
eigenvectors of the generalized eigenvalue problem
\beq{SNeigenEq}
\S\vb = \lambda \N\vb,
\eeq
sorted from highest to lowest eigenvalue $\lambda$
and normalized so that $\vb\dag\N\vb=\I$.
This implies that
\beq{SNexpecEq}
\expec{y_i y_j}=\delta_{ij}(1+\lambda_i),
\eeq
which means that the transformed data values $\y$ have the desirable
property of being
{\it statistically orthogonal}, {\ie}, 
uncorrelated.
In the approximation that the distribution function
of $\x$ is a multivariate Gaussian, this 
also implies that they are statistically independent ---
then $\y$ is merely a vector of independent
Gaussian random variables. 
Moreover, since the eigenmodes diagonalize both $\S$ and $\N$ 
simultaneously, \eq{SNeigenEq} shows that the eigenvalues $\lambda_i$ 
can be interpreted as a signal-to-noise ratio $\S/\N$.
Since \eq{SNexpecEq} shows that the quantity $y_i^2-1$ on average 
equals this signal-to-noise ratio, it is a useful band power
estimator when normalized so that its window function integrates to unity.
The window function
is given by \eq{RankOneWinEq} 
with $\psi_i$ replaced by the continuous KL mode defined by 
\beq{ContinuumModeDefEq}
\psi^\prime_i(\r)\equiv \sum_{j=1}^N \B_{ij}\psi_j(\r),
\eeq
since $y_i=\int[n(\r)/\nbar(\r)-1]\psi^\prime_i(\r)d^3r$. 
For the volume-limited case, the noise power is simply $1/\nbar$, so 
the correctly normalized and bias-corrected KL band power estimators
are simply 
``signal $=$ noise $\times$ signal-to-noise'', {\ie}, 
\beq{KLqDefEq}
q_i\equiv {y_i^2-1\over\nbar}; 
\eeq
compare this with \eq{PtDefEq}. 
Since the matrix $\B$ is invertible, the final data set $\y$ clearly retains
all the information that was present in $\x$.
In summary, the KL transformation partitions the information content of 
the original data set $\x$ into $N$ chunks that are
\begin{enumerate}
\item mutually exclusive (independent),
\item collectively exhaustive (jointly retaining all the information), and
\item sorted from best to worst in terms of their information content.
\end{enumerate}
Typically, most of the KL coefficients $y_i$ have a signal-to-noise 
ratio $\lambda\ll 1$, so that the bulk of the 
cosmological information is retained in the first
$N'$ coefficients, $N'\ll N$, which is why the KL method is 
often 
referred to and used as {\it data compression}.
One can thus throw away all but the first $N'$ numbers $y_i$
without any appreciable information loss, and this 
compressed data set will still satisfy properties 1 and 3 exactly, and
2 to a good approximation. 


Bunn (1995) and 
VS96 have pointed out that the S/N-coefficients $\y$ are useful for
power spectrum estimation
since as long as the galaxy
survey probes only scales smaller than the peak in the power 
spectrum (cf., the discussion in \Sec{powerpeakSec}),
the first $N'$ window functions $W_i$ have the following two
properties: 
\begin{enumerate}
\item They are narrow in $k$-space,
\item As $i$ increases from $1$ to $N'$, they probe 
all the scales accurately measured by the survey, 
from largest to smallest.
\end{enumerate}
Since they are also uncorrelated, 
these power spectrum estimators therefore have all desirable
properties
that one may wish for: they distill the cosmological 
information
content of the data set into a set of mutually exclusive and 
collectively exhaustive chunks, which correspond to the band 
powers
in a set of narrow bands.
In the approximation that $\y$ has a Gaussian distribution, the probability
distributions of the power estimates $q_i$ are simple:
they are independent $\chi^2$ distributions with one degree of 
freedom.\footnote{To prevent 
power spectrum plots from becoming too cluttered with
points with large error bars, it is convenient to combine neighboring
band power estimates with a weighted average --- these broader band
powers will of course still be uncorrelated since all the $q_i$ are.
}

\subsubsection{General KL modes}

\label{GeneralKLsec}

Let us write $\C=\theta_i\S+\N$,
where the power spectrum normalization parameter 
$\theta_i=1$ in the fiducial model.
Since $\S=\C,_i\equiv\partial\C/\partial\theta_i$, 
\eq{SNeigenEq} can be rewritten as (TTH)
\beq{EigenEq2}
\C,_i\vb = \lambda'\C\vb,
\eeq
where $\lambda'=\lambda/(1+\lambda)$ or, equivalently,
$\lambda=\lambda'/(1-\lambda')$.
From now on, we will normalize the eigenvectors so that
$\vb\dag\C\vb=1$ instead of $\vb\dag\N\vb=1$, since this is more
convenient throughout the rest of the paper. The matrix elements
$\B_{ij}$ are thus a factor $(1+\lambda_i)^{1/2}$ smaller 
than in the previous subsection.
As shown in TTH, solving the eigenvalue \eq{EigenEq2} is useful
for {\it any} parameter $\theta_i$ on which $\C$ depends, even those
which do not affect only the power spectrum (e.g., redshift-space
distortions, cf., Appendix C), 
and the signal-to-noise eigenmode method discussed above is just the
special case where 
$\theta_i$ is the power normalization. The three properties
listed above continue to hold in the general case, 
and the ``information content'' in item 3 above now refers to the
information about the parameter $\theta_i$.

The KL method is a very general data analysis tool.
Note that the eigenmodes continue to be mutually exclusive and collectively
exhaustive if we replace $\C,_i$
by {\it any} symmetric matrix $\M$ in \eq{EigenEq2}. To 
ensure that the KL modes give narrow window functions ranging from
small to large $k$, one can therefore choose $\M$ to be the signal
covariance matrix  
$\S$ that would arise from some monotonically decreasing 
fiducial power spectrum, for instance $P(k)\propto k^{-3}$. 
In that case, the modes become sorted by the ratio of ``signal'' in the {\it
fiducial\/} power spectrum to noise; as the former is monotonic, the
modes are sorted by scale. 
We discuss this point further in
\Sec{powerpeakSec}.

\subsubsection{KL modes are asymptotically pixelization-independent}

The pixelizations listed in \sec{PixelizationSec}
were all in some sense arbitrary, and generally somewhat redundant.
The KL modes can eliminate this arbitrariness, as mentioned in
\Sec{GuessedmodesSec}. 
If the pixelization functions $\psi_i$ formed a complete set,
spanning the space of all square-integrable functions over the survey volume,
the continuum KL modes defined by \eq{ContinuumModeDefEq}
would be some smooth functions $\psi^\prime_i(\r)$ that were
independent of the pixelization used to compute them and depended only 
on $P(k)$, $\nbar(\r)$ and the geometry of the survey.
In practice, the functions $\psi_i$ do not of course form a complete set,
since one is limited to a finite number of pixels, but
the continuum KL modes $\psi^\prime_i(\r)$ can nonetheless be computed numerically.
By choosing a pixelization that can resolve all features down to
some scale $R$, \eq{ContinuumModeDefEq} will accurately approximate
all continuum modes whose window functions $W_i(k)$ probe only scales 
$k^{-1}\gg R$.
If the number of pixels are increased further to probe smaller scales, 
the first $N'$ modes remain stable against this perturbation:
the first $N'$ eigenvalues and eigenvectors would only change
by a small amount (Szalay \& Vogeley 1997). The new modes, on the
other hand, would represent the  
small scale noise probed by the new pixel scale. 
By using the functions $\psi^\prime_i(\r)$ to pixelize the data, $i=1,...,N$,
one can thus make sure
that all of the cosmological signal down to the scale probed by
$\psi^\prime_N$ is retained.

\subsubsection{Using KL modes for trouble spotting}
\label{TroubleSpottingSec}

If the assumed power spectrum model is correct, the KL coefficients $y_i$ 
defined by \eq{yDefEq} will be independent Gaussian random
variables with zero mean and unit variance.
This offers an efficient way of testing whether the data are 
inconsistent with this model. 
The detection of, say, a $6-\sigma$ outlier ($|y_i|>6$) would provide
strong evidence that there is either more 
variance on the scale probed by the $\ith$ mode than the 
fiducial power spectrum assumed, or that the probability distribution
is strongly non-Gaussian on that scale.  
Even if it goes undetected, an 
incorrect assumed model does not
bias the estimate of the power spectrum, as discussed
below in \sec{RedHerringSec}.

\subsubsection{Using KL modes for linear filtering}

\label{KLfilteringSec}

The KL eigenmodes have an additional use. The process of throwing away the 
eigenmodes with low signal-to-noise ratios 
splits the space of all possible density fields given the data into
two --- one subspace 
that mostly contains noise, and one that is dominated by our
generalized signal. The two are statistically orthogonal to one another. 
The expansion of a given dataset over the signal subspace will 
substantially reduce the noise, thus representing a useful linear
filtering of the data (cf., Seljak 1997). 
Since it maximizes the signal for a given
number of included modes, the KL transform is sometimes referred to 
as ``optimal subspace filtering''. 
More generally, let us define a filtered data set
\beq{FilteringEq}
x'_i\equiv\sum_{j=1}^N  (\C\B\dag)_{ij} w_j y_j
\eeq
for some weights $w_j$. 
The generalized eigenvector orthogonality relation gives
$\B\C\B\dag=\I$ (cf., VS96, TTH), which implies that
$\C^{-1}=\B\dag\B$ and $\C\B\dag\B=\I$.
Hence \eq{yDefEq} gives $\C\B\dag\y=\x$, which implies that we 
recover the original data set 
($\x'=\x$) in \eq{FilteringEq} if we choose the weights $w_j=1$.
Another simple example is the
optimal subspace filtering mentioned above, which corresponds to the 
choice $w_j=1$ for $j\le N'$, $w_j=0$ otherwise.
Finally, it is easy to show that the choice
$w_j=\lambda'_j$ gives {\it Wiener filtering}, which is defined by
$\x'=\S\C^{-1}\x$. In other words, Wiener filtering becomes diagonal
in the KL basis, since it diagonalizes $\S$ and $\C$ simultaneously.
Indeed, Wiener filtering is but one of many linear filters
that are straightforward to implement in the KL basis.  

\subsection{Quadratic method}

\label{QuadraticSec}

\label{QuadSec}

In both traditional methods and the pixelized 
KL technique, the power spectrum estimates
$q_i$ are some quadratic functions of the observed density
field $n(\r)$, {\ie}, of the form given by \eq{DirectQuadEq}.
Hamilton (1997a) adopted a more ambitious approach, and 
computed
the unbiased quadratic power estimators that have minimal
variance, using a series expansion. 
T97 subsequently showed that these estimators are
unbeatable; their Fisher information matrix is identical to that of
the raw data, so no non-quadratic unbiased estimators 
can give smaller variance. 
Moreover, they can be computed 
without recourse to the numerically cumbersome series 
expansion of Hamilton (1997a) (c.f. T97 and Knox, Bond, \& Jaffe 1997 for
applications to CMB observations). 
Here we show how this method can be applied to galaxy surveys, and 
its relation to the KL method. Just as for the KL technique, a fiducial 
power spectrum is assumed.

We will parameterize the power spectrum as a piecewise constant function 
with $N'$ ``steps'' of height
$p_i$, which we term the {\bf band powers}.  Thus $P(k) = p_i\ {\rm
for}\ k_i \le k < k_{i+1}$, where
\beq{BandDefEq}
0=k_1<k_2<...<k_{N'+1}=\infty,
\eeq
and group them into an $N'$-dimensional vector $\p$.  
For the method to be strictly lossless, these bands should
be chosen to be quite narrow compared to the scales on which the
power spectrum varies. 
One first computes a compressed data vector $\q$ 
whose $N'$ elements are quadratic functions of the data set 
$\x$.
These are defined by
\beq{yDefEq2}
q_i\equiv {1\over 2}\z\dag\C_{,i}\z,
\eeq
where the vector $\z$ is given by
\beq{zDefEq}
\z\equiv\C^{-1}\x,
\eeq
and the matrix $\C_{,i}$ is defined by
\beq{CpDefEq}
\left(\C_{,i}\right)_{ab}\equiv\int_{k_i<|\k|<k_{i+1}}
\psih_a(\vk)\psih_b(\vk)^*\dV. 
\eeq
That is, $\C_{,i}$ is the derivative of
the covariance matrix $\C$ (\eq{xCovEq}) with respect to the normalization 
of the $\ith$ band, in the limit of narrow bands. 
Rewriting this as 
\beq{EmatrixDefEq}
q_i\equiv{1\over 2}\x^\dagger\E_i\x,
\eeq
where
\beq{QuadMethodEq}
\E_i\equiv\C^{-1}\C,_i\C^{-1},
\eeq
we see that the matrix $\E_i$ is simply 
a discrete version of the pair weighting 
function $E_i(\r,\r')$ in \eq{DirectQuadEq}.  Note that it is not
separable, \ie, it has rank greater than one.

For the Gaussian case, the 
Fisher information matrix for $\p$ defined by \eq{FisherDefEq} 
reduces to (VS96, TTH)
\beq{GaussFisherEq}
\F_{ij} = {1\over 2}\tr\left[\C^{-1}\C_{,i}\C^{-
1}\C_{,j}\right],
\eeq
and T97 shows that both the mean and the covariance of $\q$ 
are given in terms of $\F$:
\beqa{qExpecEq}
\expec{\q}&=&\F\p,\\
\label{qCovarEq}
\expec{\q\q^t}-\expec{\q}\expec{\q}^t &=&\F.
\eeqa
This means that $\F^{-1}\q$ is an optimal estimator
of $\p$, since its covariance matrix is precisely the inverse 
of the
Fisher matrix. Moreover, as shown in T97, compressing the 
data set
$\x$ into the coefficients $\q$ for some sufficiently narrow 
bands
is a strictly lossless procedure, retaining all the 
information about
those cosmological parameters that affect galaxy clustering 
through
the power spectrum alone.
\Eq{qExpecEq} shows that, in terms of the band powers, 
the window functions $W$ of \eq{GenQuadWinEq} for the
coefficients $q_i$  
are simply proportional to the rows of the ubiquitous Fisher 
matrix.

The coefficients $\F^{-1}\q$ tend to be both correlated 
and noisy, and therefore it is better to work with the
transformed coefficients defined by 
\beq{UncorrQuadEstEq}
\pest\equiv\F^{-1/2}\q,
\eeq
renormalized so that their window functions integrate to 
unity.
Here $\F^{1/2}$ denotes the symmetric matrix whose square 
is $\F$
-- it is readily computed by diagonalization.
As shown by Tegmark \& Hamilton (1997), these coefficients 
are
all uncorrelated (multiply \eq{qCovarEq} on both the right
and the left by $\F^{-1/2}$ to see this), and moreover tend to be very
well-behaved  
numerically, with narrow non-negative
window functions (the rows of $\F^{1/2}$) spanning the entire range of
scales probed  
by the survey.

\section{RELATIONS BETWEEN THE METHODS}

\label{RelationSec}

\subsection{Relation between KL and quadratic method}

As we will now show, the linear and quadratic pixelized 
methods are
closely related --- the latter is simply a faster 
way of
computing the same band power estimates.

Let us optimize our KL modes to estimate not the overall 
power normalization, 
but the band power in band $i$.  As we showed in
\sec{KLfilteringSec},
\beq{BtBeq}
\C^{-1}=\B\dag\B.
\eeq
Introducing the diagonal matrix 
$\LL\equiv {\rm diag}\{\lambda'_j\}$,
\eq{EigenEq2}
implies that
\beq{BCiBeq}
\B\C_{,i}\B\dag=\LL.
\eeq
We are partially suppressing the index $i$ here for simplicity, although the
eigenvectors in $\B$ and the eigenvalues in $\LL$ of course depend on which 
power band $i$ we optimize for.
Equations\eqn{yDefEq},\eqn{BtBeq} and\eqn{BCiBeq} 
allow us to rewrite \eq{yDefEq2} as
\beqa{yEq3}
2 q_i 
&=& \x^\dagger\C^{-1}\C_{,i}\C^{-1}\x
= \x^\dagger\B\dag\B\C_{,i}\B\dag\B\x\nonumber\\
&=&\y\dag\LL\y
= \sum_j\lambda'_j y_j^2,
\eeqa
\ie, as a weighted average of the squared KL coefficients
$y_j$, with weights given by the KL eigenvalues
$\lambda'_j$.
Thus the coefficients are weighted by their inverse
variance, which means that this is the minimum-variance band power
estimator based on the squared KL coefficients, 
so after
subtracting off shot noise and normalizing correctly, the
linear (KL) and quadratic methods give exactly the same 
estimator
for the band power.  As described in \Sec{quadproconSec}, the quadratic
estimator is is simply faster to 
compute. 
A more intuitive way of understanding why the two methods
give the same result is to note that the
quadratic method was derived explicitly to be the minimum
variance estimator (T97; cf., \eq{qExpecEq}), and that
VS96 showed that the KL approach is  
guaranteed to minimize the variance as well.

\subsection{Relation between quadratic and ML methods: iteration}

\label{QuadMLrelSec}

If the quadratic method is repeated using the output 
(measured) power spectrum as the input (fiducial) power spectrum,
then this iteration will eventually converge to the power spectrum that
would be obtained with the brute force maximum-likelihood method
(Bond, Jaffe, \& Knox 1997). This is because the quadratic method can be derived by 
expanding the logarithmic likelihood $\ln f$ 
to second order around the 
fiducial point and maximizing it. The iteration thus seeks the maximum
of $\ln f$ by repeatedly approximating it with a parabola. This is 
essentially the maximum-gradient method for maximization, which
is known to have excellent convergence properties. In the one-dimensional
case, this is equivalent to finding the zero of $(\ln f)'$
by repeated linear approximations. This is simply the Newton-Raphson
root-finding method, known to converge exponentially fast and 
asymptotically double the number of correct decimals in each iteration.

Although the ML-result can be computed by iterating the 
quadratic method, it should be stressed that 
this is not necessarily a good thing to do.
If the measured power spectrum is noisy, so that the band power 
uncertainties do not satisfy $\Delta P\ll P$, then iteration
can produce misleadingly small error bars,
as the following example illustrates.
Suppose sample variance or a shot noise fluctuation 
makes a band power measurement ten times smaller than the true value
and this is used as the new fiducial band power $P$ (prior). Then 
\eq{FKPdPEq} shows that the nominal $\Delta P$ from sample variance
will be ten times too small as well.
If iteration is nonetheless desired, it is crucial not to let the prior
over-fit the data. A better approach than simple iteration is therefore
to use as prior a fit to the measured $P(k)$ with a smooth parameterized 
fitting function, and keep adding fitting parameters until the 
the value of $\chi^2$ per degree of freedom drops below unity (Tegmark
1997a; Seljak 1997).

\subsection{Relation between quadratic and FKP methods: the small scale limit}

\label{SmallScaleSec}

For a traditional method with a volume weighting function $\phi(\r)$,
we define a quantity $L$ implicitly by 
\beq{LdefEq}
\int_{kL<1} |\phih(\k)|^2\dV = {1\over 2},
\eeq
\ie, $L^{-1}$ is the radius of a sphere in Fourier space containing half 
of the $\k=\vzero$ window function.
For simple volumes such as a pencil beams, slices or spheres, 
$L$ is of the order of the width of the survey in its {\it narrowest} 
direction.
We will now show that in the small-scale limit where $k^{-1}\ll L$,
the quadratic method reduces to the FKP method, 
which implies that the latter is lossless and hence unbeatable
for measuring the power spectrum on the smallest scales,
as was pointed out by Hamilton (1997a).

\subsubsection{Derivation of the FKP method} 

\label{FKPsec}

The FKP method (Feldman, Kaiser \& Peacock 1994) is the traditional
method described in \sec{TraditionalSec}, using the 
volume weighting of \eq{FKPeq}. The rank one power estimates $|x_a^2|$
are averaged with
equal weights\footnote{
An intuitive way to understand why all directions in 
$k$-space
should receive equal weight when $\Delta k\gg L^{-1}$ is to 
note that in this limit, the number of coherence volumes that
fit into a given solid angle in the shell is independent 
of its shape (and hence independent of 
direction in $k$-space),
being merely the ratio of the shell subvolume to 
the coherence volume.
}
for all modes in a spherical shell $|\k|=k_a$,
so integrating \eq{RankOneEq} over this shell
using \eq{FourierPixelizationEq} for the pixels,
we see that this corresponds to 
the pair weighting
\beq{SimplePtEq}
E_\i(\r,\r') = \phi(\r)\phi(\r')j_0(k_i|\r-\r'|),
\eeq
where $j_0(x)\equiv\sin(x)/x$.
This holds for any choice of volume weighting function $\phi$.
The specific FKP choice given by \eq{FKPeq} is derived by minimizing 
the variance of the corresponding power estimators $q_\i$.
This involves a number of approximations.
We summarize the derivation here, to clarify why the FKP weighting is
optimal only on small scales.

Substituting equations\eqn{xDefEq} and\eqn{FourierPixelizationEq}
into equations\eqn{NdefEq} and\eqn{SdefEq}, we obtain 
the pixel covariance matrix
\beqa{TradCovEq}
\C_{ab}
&=&
\int\phih(\k-\k_a) \phih(\k-\k_b)^* P(k)\dV\nonumber\\
&+& 
\int e^{i(\k_a-\k_b)\cdot\r} {\phi(\r)^2\over\nbar(\r)}
d^3r.
\eeqa
The function $\phih(\k)$ roughly falls off on a 
scale $L^{-1}$.
As long as $L^{-1}\ll k$ and 
the power spectrum $P(k)$ is a smooth function, $P(k)$
will therefore be almost constant where the first 
integrand is non-negligible (\ie, where $|\k-\k_a|\simlt L^{-1}$
and $|\k-\k_b|\simlt L^{-1}$) and 
can be approximately factored out of the $k$-integral. 
Using the convolution theorem, this yields
\beq{SimplerCorrEq}
\C_{ab}\approx\int e^{i(\k_a-\k_b)\cdot\r}
\phi(\r)^2\left[P+{1\over\nbar(\r)}\right]d^3 r,
\eeq
where $P\equiv P([k_a+k_b]/2)$.
This shows that $\C_{ab}\approx 0$ if 
$|\k_a-\k_b|\gg L^{-1}$, so power estimates separated by
much more than $\Delta k=L^{-1}$ are essentially uncorrelated. 
Conversely, power estimates from nearby shells with
$|\k_a-\k_b|\ll L^{-1}$ are almost perfectly correlated and 
therefore redundant. The FKP method therefore 
averages the power estimates $|x_a|^2$ in thicker shells 
$|\k_a|\in[k_i,k_{i+1}]$ as in \eq{BandDefEq},
whose widths satisfy 
$L^{-1}\ll |k_{i+1}-k_i|\ll k$, giving approximately uncorrelated
power estimates $q_\i$. 
This approximation clearly only holds for small scales, $k^{-1}\ll L$.
We denote the volume of the $\ith$ shell 
\beq{VsDefEq}
V_s\equiv {4\over 3}\pi\left(k_{i+1}^3-k_i^3\right)/(2\pi)^3.
\eeq
Assuming that the pixelized data $\x$ is Gaussian distributed, 
the variance of the power estimator $q_\i$ is simply given by
averaging $2|C_{ab}|^2$ over the shell, \ie,
\beq{PixelizedVarianceEq}
(\Delta q_\i)^2 = {2\over V_s^2}\int\!\int|\C_{ab}|^2 
{d^3 k_a\over(2\pi)^3} {d^3 k_b\over(2\pi)^3},
\eeq
where both integrals are to be taken over the $\ith$ shell.
Substituting \eq{SimplerCorrEq},
and using the fact that
$L^{-1}\ll|\k_a-\k_b|\ll k$, one of these integrals simply
produces a factor $V_s$. Applying Parseval's theorem to
the result and using \eq{RankOneNormEq} to normalize\footnote{
Since $kL\gg 1$, \eq{RankOneNormEq} does not need the 
integral constraint correction.}
finally leaves us with the approximation
\beq{dPApproxEq}
\left({\Delta q_\i\over\expec{q_\i}}\right)^2
\approx
2{\int\phi(\r)^4
\left[1+{1\over\nbar(\r)P}\right]^2 d^3 r\over
V_s\left[\int\phi(\r)^2 d^3 r\right]^2}.
\eeq
This holds for any weighting function $\phi$. The FKP choice of $\phi$ 
is derived by minimizing
this approximate expression for the variance. 
Since it is left unchanged if we multiply $\phi$ by a 
constant,
we can for simplicity impose the normalization constraint
$\int\phi^2 d^3r=1$. Minimizing the numerator of 
\eq{dPApproxEq} with a Lagrange
multiplier for this constraint now gives the FKP weighting of
\eq{FKPeq}.
Substituting this back into \eq{dPApproxEq} and omitting the 
power band index $i$ for simplicity finally yields
\beq{FKPdPEq}
\left({\Delta P\over P}\right)^2 \equiv 
\left({\Delta q\over\expec{q}}\right)^2\approx {2\over V_s\Veff},
\eeq
where 
\beq{VeffDefEq}
\Veff(k)\equiv\int\left[{\nbar(\r)P(k)\over 1+\nbar(\r)P(k)}\right]^2 d^3 r
\eeq
can be interpreted as the {\it effective volume} probed,
since the integrand is of order unity where one is signal dominated
($P\gg 1/\nbar$) and $\sim 0$ where one is noise dominated. 
For a volume-limited survey with 
spatial volume $V$ and constant $\nbar$, the FKP prescription  
weights all galaxies equally, and we simply have 
$\Veff=[1+1/\nbar P]^{-2}V$.
An intuitive way to understand \eq{FKPdPEq} is to note
that the nearby Fourier amplitudes $x_i$ are correlated 
over a {\it coherence volume} $V_c\sim 1/V$.
Thus as long as shot noise is not dominant, 
$\Delta P/P\sim \sqrt{2/\cal N}$, where $\cal N$ is the 
number of approximately uncorrelated volumes $V_c$ that fit into the
shell volume $V_s$.
In conclusion, we have shown that for traditional methods, 
the FKP volume weighting of \eq{FKPeq} is optimal 
if and only if we limit ourselves to small scales, $k^{-1}\ll L$.

\subsubsection{The small-scale limit of the quadratic method}

\label{QuadLimitSec}

It is often convenient to work in the continuum limit where 
a discrete index $i$ is replaced by a continuous variable such 
as $\r$ or $\k$. Vectors $a_i$ and matrices $\A_{ij}$ then 
correspond to functions of one and two variables, such as 
$a(\r)$ and $A(\r,\r')$. Since all sums get replaced by integrals
in this limit, units frequently differ from the discrete case
\footnote{
Since $\vb=\A\x$ means $b_i=\sum_j\A_{ij}x_j$  in the discrete case  
and $b(\r)=\int\A(\r,\r')x(\r')d^3r'$ in the continuous case, we see
that matrix multiplication introduces units of the volume element,
in this example $d^3r$. Thus the continuous analog of the identity 
matrix, $\I(\r,\r')\equiv\dirac(\r-\r')$, is its own inverse,
since $(\I^2)(\r,\r')\equiv\int \I(\r,\r'')\I(\r'',\r')d^3r''=\I(\r,\r')$, even though
it is not dimensionless --- the delta function $\dirac$
has units of inverse volume.
}.

Let us take the pixelization functions to be
Dirac delta functions $\psi_i(\r)\equiv \dirac(\r-\r_i)$, corresponding
to a continuum of pixels
\beq{ContinuumPixelEq}
x(\r) = {n(\r)\over\nbar(\r)}-1,
\eeq
and choose our parameters to be the values of the 3D power spectrum:
\beq{ContinuumParameterEq}
\theta_i = P(\k_i).
\eeq
The matrices $\C$ and $\C,_i$ then reduce to 
\beq{ContinuumCeq}
\C(\r,\r') = 
\int e^{-i\k\cdot(\r-\r')}P(\k)\dV 
+ {\dirac(\r-\r')\over\nbar(\r)},
\eeq
\beq{ContinuumCpEq}
\C,_i(\r,\r') = {\partial\C(\r,\r')\over\partial P(\k_i)}
= {1\over(2\pi)^3}e^{-i\k_i\cdot(\r-\r')},
\eeq
since $P(\k)=\int P(\k_i)\dirac(\k-\k_i)d^3\k_i$
implies that $\partial  P(\k)/\partial  P(\k_i)=(2\pi)^3\dirac(\k-\k_i)$.
As we saw in \sec{FKPsec},
the small-scale limit corresponds to neglecting the $k$-dependence of $P$, which 
gives $\C(\r,\r') \approx [P+1/\nbar(\r)]\dirac(\r-\r')$ and 
\beq{ContinuumCinvEq}
\C^{-1}(\r,\r') = \left[P+{1\over\nbar(\r)}\right]^{-1}\dirac(\r-\r')
= \phi(\r)\dirac(\r-\r'),
\eeq
where $\phi$ is the FKP weighting function of \eq{FKPeq} 
normalized so that $\phi=\nbar/(1+\nbar P)$.

We will now rederive the FKP results with much less effort than in the
previous section, by simply using the quadratic method formulas.
Substituting equations\eqn{ContinuumCpEq} and\eqn{ContinuumCinvEq}
into \eq{yDefEq2} shows that the quadratic estimators $\q$ are given by
\beq{yDefEq3}
q_i\equiv {1\over 2}|\zh(\k_i)|^2,
\eeq
where the function $z(\r)\equiv\phi(\r)x(\r)$,
which apart from the factor of $1/2$ are exactly the FKP 
estimators of $P(\k)$.
Calculating the Fisher matrix by substituting 
equations\eqn{ContinuumCpEq} and\eqn{ContinuumCinvEq}
into \eq{GaussFisherEq} gives
\beqa{FKPfisherEq}
\F_{ij} 
&=&{1\over 2}\tr\left[\C^{-1}\C,_i\C^{-1}\C,_j\right]\nonumber\\
&\approx&{1\over 2}\int\phi(\r)e^{-i\k_i\cdot(\r-\r')}
\phi(\r')e^{-i\k_j\cdot(\r'-\r)}d^3r\>d^3r'\nonumber\\
&=&{1\over 2}|\phih(\k_i-\k_j)|^2.
\eeqa
Equation\eqn{qExpecEq} now gives
\beq{FKPqExpecEq}
\expec{q_i}=\int\F_{ij}P(\k_j){d^3k_j\over(2\pi)^3},
\eeq
which shows that the 
Fisher matrix is simply the 3D window function.
To make the window function of the power estimate
$q_i$ integrate to unity, we need to divide it by the quantity
\beq{N2defEq}
\int\F_{ij}{d^3k_j\over(2\pi)^3} = {\Veff\over 2 P^2},
\eeq
where we have used \eq{FKPfisherEq}, Parseval's theorem,
\eq{FKPeq} and \eq{VeffDefEq} in the last step.
Using \eq{qCovarEq}, this shows that
\beq{SingleFKPmodeErrorEq}
{\Delta q_i\over\expec{q_i}}\approx \sqrt{2}
{\int\phi(\r)d^3r\over\int\phi(\r)^2 d^3r}
=\sqrt{2}\left[1+{1\over\nbar P}\right],
\eeq
where the last equal sign only holds for the volume limited
case where $\nbar$ is constant.
Alternatively, averaging the power estimates $q_i$ over a 
shell in $\k$-space
as in the previous section, \eq{qCovarEq} reproduces
the FKP error formula of \eq{FKPdPEq}.
In summary, we have shown that the FKP method becomes
lossless for measuring the power on small scales, 
since it equals the optimal quadratic method in this limit.

%
%

\section{SYSTEMATIC PROBLEMS --- THE SELECTION FUNCTION AND
EXTINCTION}
\label{SystematicsSec}

We have discussed the importance of the integral constraint in
\Sec{TraditionalSec}.  Here we discuss this issue in the
context of pixelized methods, and generalize it to a whole host of
systematic problems, including errors in our assumed selection
function $\bar n(\r)$ and our assumed extinction map. 


\subsection{What is the integral constraint?}
\label{ConstraintDiscussionSec}

If we knew the selection function $\nbar(\r)$ {\it a priori}, 
before counting the galaxies in our survey, 
we would be able to measure 
the power on the scale of the survey. 
Our power spectrum estimate would essentially be the square of the ratio of 
the observed and expected number of galaxies in our sample.
Of course, we do not know $\nbar$ {\it a priori}, so we use the
galaxies themselves to normalize the selection function.
Thus the measured density fluctuation automatically 
vanishes on the scale of the survey, and 
naive
application of
any of the power spectrum estimation methods we have described
will falsely indicate that $P(k)\to 0$ as $k\to 0$, regardless of
the behavior of the true power spectrum on large scales
(Peacock \& Nicholson 1991). See also Tadros \& Efstathiou (1996).
\Eq{GenQuadExpecEq} tells us that apart from the noise 
bias $b_\i$, there is an additional term 
$W_\i(\vzero)$ 
that must be 
subtracted off to make $q_\i$ an unbiased power spectrum estimator.
Let us assume that we know the shape of the selection 
function but not its normalization. To reflect this, we write the true
selection function as 
\beq{etaDefEq}
\nbar(\r)=\nnorm\nbar_0(\r),
\eeq
where $\nbar_0$ is our guessed selection function, and $\nnorm$ 
is an unknown normalization constant, and find that our noise bias
corrected power estimator will have a residual bias 
$(\nnorm-1)^2 W_\i(\vzero)$.
Since we do not know the exact value of 
$\nnorm$, our only way to eliminate this bias is to
require that $W_\i(\vzero)=0$, \ie, by 
requiring the window function to vanish at $\k=\vzero$. 
This was first explicitly pointed out by Fisher {\etal} (1993).

\subsection{Its relation to extinction and other systematic problems}

Since requiring $W_\i(\vzero)=0$ eliminates the integral 
constraint problem, the trouble is confined 
to the $\k=\vzero$ mode.
If $W_\i(\vzero)=0$, then the power estimate $q_\i$ is clearly completely 
independent of $P(0)$, the fluctuations in this untrustworthy mode.
The essence of our approach is therefore the following:

\begin{itemize}
\item 
{\it  
We have a systematic problem 
with a certain mode, and can 
immunize our results from this problem by making them independent 
of this untrustworthy mode. 
}
\end{itemize}
When phrased in this way, it is clear that this approach can
be applied to a variety of other systematic problems as well.
For instance, incorrectly modeled extinction adds excess power 
in the form of purely angular modes (density fluctuations that have no radial
component, {\ie}, being perpendicular to the line of sight).
It might therefore be desirable 
to make the results independent of all purely angular modes on 
the relevant scales or, in a less ambitious approach, at least 
independent of those modes whose shape coincide with known 
dust templates. 
Similarly, mis-estimating the {\it shape} of the radial selection function
(of $\nbar_0$ in \eq{etaDefEq})
pollutes certain purely radial modes which one may wish to discard.

\subsection{How to eliminate untrustworthy modes with pixelized methods}

In this section, we show how the power spectrum estimate from a pixelized  method
can be made insensitive to the type of systematic errors described above.


Let us parameterize the true selection function $\nbar$ as
\beq{ParametrizedNbarEq}
\nbar(\r)=\sum_{j=1}^M \nnorm_j\nbar_j(\r),
\eeq
where $\nbar_j$ are known functions and
the parameters $\nnorm_j$, which we group into
an $M$-dimensional vector $\vnnorm$, are 
{\it a priori} unknown. 
Imagine for example a simple case where $M=3$, 
$\nbar_1$ is our best guess for a purely radial
selection function based on a Schechter luminosity function, 
$\nbar_2$ is a (purely angular) dust template, 
and $\nbar_3$ gives the effect of an infinitesimal error in the
estimate of the characteristic luminosity $L_*$:
$\nbar_3(\r) = \partial \nbar_1(\r)/\partial L_*$. 
Let $\nbar_0$ denote some
{\it a priori} estimate of $\nbar$.
Defining the ``uncorrected" pixels as
\beq{xpDefEq}
x'_i \equiv\int{n(\r)\over\nbar_0(\r)}\psi_i(\r) d^3 r,
\eeq
we find that 
\beq{xpExpecEq}
\expec{\x'}=\Z\vnnorm,
\eeq
where the $N\times M$  matrix $\Z$ of untrustworthy modes
is defined by
\beq{ZdefEq}
\Z_{ij}\equiv\int{\nbar_j(\r)\over\nbar_0(\r)}\psi_i(\r) d^3 r.
\eeq
This means that in general, the data set $\expec{\x'}\neq 0$,
so the uncorrected data set does not satisfy \eq{xExpecEq}.

We show how to solve this problem in Appendix B. In short, 
one replaces $\x$ by a cleaned data set $\PP\x$, where $\PP$ 
is a matrix that satisfies $\PP\U = \vzero$ and thus projects 
out the untrustworthy modes.
We find that the best choice is
$\PP=\I - \Z(\Z\dag\C^{-1}\Z)^{-1}\Z\dag\C^{-1}$.
Although the covariance matrix of the cleaned data set is not invertible,
we find that with this choice of $\PP$, the quadratic method remains
strictly optimal if we simply use $\C$ from the 
uncleaned data in \eq{zDefEq}. We also show that the integral constraint
correction given by equations\eqn{NormDefEq} and\eqn{sshotDefEq}
corresponds to this optimal method in the small scale limit.

\section{PROS AND CONS OF THE METHODS}

\label{ProsConsSec}

Above we have presented all methods for galaxy power spectrum estimation
that have proposed in the literature, as well as the new pixelized quadratic 
technique and various extensions, and showed how they are related to one another.
We will now discuss their relative merits at some length. It will become 
clear that they are highly complementary, and we summarize the pros and
cons of each method in Table~1. 
We will return to this in the discussion section, where we describe
how the traditional, KL and quadratic methods can be used in concert
to produce a data analysis pipeline having all the properties
on our wish list from Section 2.

\bigskip

\noindent\RatingTable
\bigskip

\subsection{Pros and cons of the direct Fourier method}

Unlike the pixelized methods, the traditional Fourier method
uses the exact galaxy positions and thus retains all the small-scale 
information about $P(k)$.

As discussed in \sec{SmallScaleSec}, this method becomes lossless in the limit 
$k L\to\infty$ when the FKP choice of $\phi$, \eq{FKPeq}, is used. 
In addition, 
choosing the shell widths $\Delta k\gg L^{-1}$ guarantees 
that the errors in the band power estimates $\Pt_i$ will be approximately 
uncorrelated. This means that on small
scales, say $k L \ll 10\%$, this method satisfies the first 
three criteria on our wish list in \sec{WishListSec}.

However, these advantages no longer hold when measuring
power on scales comparable to the size of the survey.
VS96 review  
problems with the direct Fourier approach that occur
unless $k L\ll 1$.  They fall into the two categories
described below.
 
\subsubsection{The direct Fourier method destroys information}

Hamilton (1997b) has shown that a strictly optimal direct 
summation
method can be derived {\it in principle}, 
in terms of a series expansion,
but this is unfortunately extremely burdensome numerically 
except for scales much smaller than the survey size, 
away from the boundaries, 
where it approximates equations\eqn{FKPeq} and\eqn{T95eq}.
The optimal galaxy pair weighting is not separable 
(in the sense of \eq{RankOneEq})
and thus cannot be expressed in terms of a 
volume weighting function $\phi$ as above.
Consequently, no direct Fourier methods are lossless
except on small scales. 

\subsubsection{The method is complicated and computationally slow for large scales}

It is important to note that the Fourier transformation calculation 
takes only a negligible amount of time in a power spectrum analysis
--- the lion's share of the work involves computing 
the mean corrections $W_i(\vzero)$, 
the shot noise correction $b_i$, 
the normalization factor $A_i$ and the covariance matrix of 
the power estimates.
This becomes numerically cumbersome on the largest 
scales, when $kL\sim 1$.
This is because both smearing from the window function and 
the effect of the integral constraint become important and 
must be accurately computed in this regime. Even though 
the relevant integrals can be greatly accelerated with FFT's,
the calculations are not only much more complicated and 
obscure than the simple linear algebra of the pixelized methods, 
but generally substantially slower as well. The most
time-consuming step is the computation of the error
bars $\Delta P$ and the covariance between power 
estimates using an
integral constraint corrected and 
appropriately generalized version of 
\eq{PixelizedVarianceEq}, since the integral in 
\eq{TradCovEq} must be done
separately for each {\it pair} of grid points $(\k_i,\k_j)$ 
in Fourier space (cf., Goldberg \& Strauss 1998). 
The property of uncorrelated errors
is clearly lost, making it quite difficult to compute the optimal 
weights for averaging the power estimates in $\k$-space into
power bands in $k$-space (T95, VS96).
In short, when comparing the direct Fourier method
to the KL and quadratic methods on the largest scales, 
it is more complicated, uses more CPU time and produces an inferior result.

\subsection{Pros and cons of the brute force method}

The pixelized brute force method has been applied 
to galaxy survey analysis using expansions in spherical harmonics
(Fisher {\etal} 1994; Heavens \& Taylor 1995; Ballinger, Heavens \&
Taylor 1996). 

It is arguably the simplest of all methods, both conceptually
and in implementation.  Moreover, it can be shown to be lossless in
the limit of large  data sets, 
thus giving minimal error bars.
An important disadvantage is that it is slow. The slowest 
step in 
evaluating the likelihood function $f$ is to compute 
$\det\C$, for which the
CPU time required scales as $N^3$, and $f$ is evaluated at a 
large number of
grid points in the multi-dimensional parameter space.
This is why it is so useful if $N$, the number of pixels, 
can be reduced by a lossless data compression scheme before performing 
the likelihood analysis, throwing away noise and keeping the signal.
As we described above, this is exactly what the 
Karhunen-Lo\`eve and quadratic methods do.
In the KL scheme,
the compressed data set $\y$ was a linear function of $\x$ (of 
the form $\y=\B\x$ for
some matrix $\B$), and in the quadratic scheme $\q$ was a 
quadratic function
of $\x$, of the form $q_i=\x^\dagger\E_\i\x/2$ for some matrices 
$\E_\i$.

A second disadvantage is that the maximum likelihood parameter
estimate $\vtheta_{ml}$ (defined in \sec{BruteSec}) is 
such a complicated function of $\x$ that
we cannot calculate its statistical properties analytically. 
As we saw above, 
both the linear and quadratic schemes allow us to write down 
power spectrum estimators
in closed form (equations\eqn{KLqDefEq} and\eqn{yDefEq2}, respectively)
This allows one to 
compute their probability
distributions exactly, 
(the $y_i$ are Gaussian and the $q_i$ are
are generalized 
$\chi^2$ distributions, often close to Gaussian), 
which 
makes these power 
estimates easy to use for parameter fitting further down the 
data analysis pipeline.

\subsection{Pros and cons of the KL method}

\subsubsection{It retains the phase information}

An important advantage of the KL method over the others in 
the table
is that the KL coefficients retain the spatial information 
about the data
(not only the Fourier amplitudes, but the corresponding 
phases as well).
This means that information on processes 
which affect the radial and angular clustering patterns
differently can be optimally probed with the 
KL method. Such effects include
\begin{enumerate}
\item Redshift space distortions (cf., Appendix C),
\item Galactic extinction (which affects only angular modes),
\item Evolution and mis-estimates of $\nbar(r)$ (which affect only radial modes).
\end{enumerate}

We have not discussed in any detail how one might include these effects in a KL
analysis.  Suffice it to say that for any physical effect that affects
the covariance matrix of the pixels, one can determine KL modes which
are optimized for parameters which describe this effect; cf.,
\sec{GeneralKLsec}.  The way to do this for redshift-space distortions
(see Appendix C) is described in more detail in TTH.  
We argue below in \sec{MultiParameterSec}
that the simple signal-to-noise
eigenmodes are appropriate general purpose modes for a KL analysis of 
any parameter which affects the covariance matrix only via the power spectrum.
On the other hand, the KL modes should be custom tailored
(using \eq{EigenEq2}) for parameters not in this category, 
such as ones causing anisotropic clustering.

Another approach is that discussed in \sec{SystematicsSec} and
Appendix B: rather than {\it
measuring\/} these systematic effects, one can project out those modes
that are sensitive to them, making the resulting dataset immune from
them.  This can be done for any pixelized method,
but is especially simple for the quadratic method.


%


\subsubsection{Pixelization is not lossless}

The ``$-$" on row 2 of Table 1 refers to the fact that 
computational constraints place an upper limit on the number
of pixels $N$ used for the eigenvalue problem, 
since the storage required scales as $N^2$ and the CPU time 
as $N^3$.
$N=10^4$ is readily handled on a high-end 1997 workstation 
(TTH), 
and new methods under development  (Szalay \& Vogeley 1997) may well
be able to increase this by an order of magnitude or more,
but since the dynamic range is 
$\sim N^{1/3}\propto$(CPU time)$^{1/9}$, 
some information
will always be lost on the very smallest scales.
Fortunately, this is not a problem in practice, since the 
complementary traditional methods
work best precisely on the smallest scales.

\subsubsection{The KL power peak problem}
\label{powerpeakSec}
For very deep surveys such as SDSS and 2dF, which 
probe scales substantially beyond the expected
peak in the power spectrum at $\sim 200h^{-1}\Mpc$, 
the KL window functions will generally 
not be narrow but double-peaked,
since fluctuations
longward of the peak have the same signal-to-noise ratio as 
certain fluctuations 
shortward of the peak and will get mixed in the corresponding 
KL modes.
One readily circumvents this degeneracy problem by performing 
a likelihood analysis 
on the KL modes, with the band powers being the parameters to 
be estimated.
However, the statistical errors on the resulting band power 
estimates will no 
longer be uncorrelated, and since this
is a nonlinear operation, they will also not have the simple 
$\chi^2$ 
distribution in general.
This is why Table 1 indicates ``$+/-$" in row 4: the ``+" 
applies 
when we use the direct approach on a volume smaller than 
$\sim 200 h^{-1}$ Mpc in size, and the ``$-$" applies when we 
use the
indirect approach on a deep data set such as SDSS or 2dF.
As was described in \sec{GeneralKLsec}, 
one can circumvent this power peak problem 
by using choosing a monotonically decreasing fiducial power
spectrum such as $P(k)\propto k^{-3}$. 
This can hardly be said to make the KL-modes
less ``optimal'', since they will still partition the
information into mutually 
exclusive and collectively exhaustive chunks. 
They simply become sorted according to a different criterion: not 
by their information content regarding the power normalization, 
but by the physical scale they probe.

\subsubsection{The KL multi-parameter complication}
\label{MultiParameterSec}

The derivation of the signal-to-noise eigenmode
method in {\eg} VS96 or TTH does 
{\it not} prove that the compressed data set 
retains the bulk of the information about 
all parameters of cosmological interest, 
but merely that it is lossless with respect to 
the overall power spectrum normalization.
In fact, this
is not a problem in practice.
TTH describe a method where one carries out a series of KL
transforms, optimizing for each parameter of interest in turn, then
pools all the resulting eigenmodes and uses singular value decomposition
to eliminate redundancy from the pool of modes. 
In addition, there is good reason to believe that 
the information about all cosmological parameters is 
nonetheless preserved even with the signal-to-noise eigenmodes
alone.
We now give a hand-waving
argument to this effect, and
describe some detailed numerical experiments that support it.
TTH computed three separate sets of KL modes 
for the CMB data set of the COBE satellite ($N=4016$), 
optimized for 
measuring three different parameters: the power spectrum
normalization $Q$, the slope $n$ and the reionization
parameter $\tau$. The $3\times 3$ Fisher matrix 
was then computed from
the three compressed data sets separately for $N' = 500$ modes
retained. 
Each set of modes retained 
virtually
all the information about their corresponding parameter, but 
the error bars on $Q$ with the $n$ or $\tau$ modes were substantially 
larger than 
their Cram\'er-Rao minimum.  That is, the $n$-modes and the 
$\tau$-modes
lost information about $Q$.  On the other hand, the 
$Q$-modes
were found to retain virtually all the information about $n$ 
and $\tau$.
Examination of their window functions revealed why. To 
obtain a large lever arm for determining the slope, 
the $n$-modes were probing mainly the largest and the 
smallest
available scales, ignoring those in the middle near the 
``pivot point".
The $\tau$-modes were ignoring the very largest scales, 
since these are unaffected by reionization and therefore 
carry no information about $\tau$.
The $Q$-modes, on the other hand, were faithfully probing the 
power
on all available scales, and therefore automatically retained 
all of the information about $n$ and $\tau$ as well, as a 
side effect.
Thus as long as the galaxy power spectrum has no sharp features 
(which the
signal-to-noise eigenmodes might potentially ignore)
we expect the standard 
KL modes optimized for 
the normalization to be close to lossless with respect to all 
cosmological parameters affecting only the power spectrum.

\subsubsection{Does KL bias the results towards our theoretical 
prejudice?}

\label{RedHerringSec}

It has been argued that the KL method 
biases the results by guessing an {\it a priori}
fiducial power spectrum when computing the pixel covariance matrix.
This claim has been extensively tested numerically (Bunn 
1995, TTH), and
found to be completely unfounded. In general, the
effect of guessing an incorrect prior model
is to leave the estimates unbiased, but with slightly larger
error bars than what is optimal. If desired, 
any dependence on the initial data can clearly be eliminated 
by iterating the KL procedure
as described in \sec{QuadMLrelSec}, 
while at the same time reducing the error bars 
close to the minimum allowed by the Fisher matrix.



\subsection{Pros and cons of the quadratic method}
\label{quadproconSec}

One advantage of this method is that its simple and 
uncorrelated 
quadratic estimators
have narrow window functions at all $k$, even beyond the peak in the 
power spectrum.
Thus it is a useful complement to the KL method on the
very largest scales, as indicated by row 4 in Table 1.

A second advantage is evident from \eq{yDefEq2}: 
Since the CPU time for multiplication of a vector by a
matrix scales as $N^2$, the time for computing 
$\z$ scales as $N^2$ as well if \eq{zDefEq} is solved
by an iterative technique such as the conjugate gradient
method (Press {\etal} 1992).  This is much faster than the KL method,
which scales as $N^3$. 
This speed increase may allow the quadratic method to be used
over a somewhat larger dynamical range than the KL method, 
extending down to smaller scales.
The quadratic method can also be used as a faster way to 
obtain the same results as the brute force ML method, 
using the iteration scheme described in \sec{QuadMLrelSec}. 

Furthermore, we saw in \sec{SystematicsSec} and Appendix B that the
quadratic method can be made immune to various sorts of systematic
errors which might plague the data. 

An important disadvantage is that, unlike the KL method, it does not 
retain any
phase information. This is a drawback 
when estimating the underlying real-space power spectrum,
since although it can measure this
directly by computing the appropriate $\C$ 
once the distortion parameter $\beta\equiv\Omega^{0.6}/b$ is 
known, the KL method or another linear approach must be used first, 
to measure $\beta$. Once cannot simply immunize the data from 
mis-estimates of $\beta$ using the formalism of Appendix B,
since $\beta$ affects virtually {\it all} the modes.
%
%
It does, however, appear possible to 
generalize the quadratic method to overcome
this limitation (Hamilton 1997, private communication).

\section{DISCUSSION \& CONCLUSIONS}
\label{DiscussionSec}

In this section, we summarize our discussion of the pros and 
cons of
the various methods, and conclude by describing an 
approach combining the strengths of all of them, illustrated 
in 
Figure~1.

We found that although the direct Fourier approach is 
both simple to 
implement and virtually lossless on scales much smaller than the
smallest dimension of the sample in question,
it has several drawbacks on larger scales: 
\begin{enumerate}
\item It loses information, giving unnecessarily noisy 
measurements.
\item It is quite tedious to implement numerically if one 
uses the exact 
expressions we have derived for the integral constraint 
correction, especially for computing the covariance.
\end{enumerate}
In contrast, the two pixelized methods are lossless on {\it 
large}
scales, but lose small-scale information because numerical 
constraints
on the number of pixels limit the dynamical range. Since they 
are 
simpler to implement as well,
they allow a more ambitious approach incorporating 
complications such as
redshift-space distortions, residual extinction and radial 
selection 
function errors (\sec{SystematicsSec}, Appendices B and C). 
The quadratic method can compute exactly the same band powers 
as the KL method, and do so  
faster (the number of operations scaling as the square rather 
than the cube
of the number of pixels), allowing more pixels and a larger 
dynamical range.
The KL method, on the other hand, is the only one which 
retains 
the phase information in which clustering anisotropies 
(differences between the
angular and radial clustering patterns) is encoded. Since 
redshift distortions and various systematic problems manifest 
themselves
in this way, the KL method is therefore a powerful complement 
to the 
quadratic method, since the former can quantify and subtract 
these systematic effects and pass the 
appropriate redshift-distortion parameter $\beta$ 
along to the latter, which can then measure the
power spectrum directly in real space as described in Appendix C.
Alternatively, the
quadratic method can be immunized from such systematic effects, 
as described in Appendix B.
The KL method is also useful 
for cosmographic purposes, where spatial information 
is everything.   Finally, the KL method also has the advantage
of greatly simplifying trouble spotting such as search for outliers
and non-Gaussian behavior. 


In conclusion, we have found that although none of the  
methods
can be made both feasible and lossless on its own, we can 
obtain a 
feasible and virtually 
lossless data analysis
pipeline satisfying our entire wish list 
by combining three of them, as outlined in Figure~1.

\begin{enumerate}
\item The power spectrum on scales $k^{-1}\simlt L/10$ is
estimated directly from the raw redshift data with the 
traditional
direct Fourier approach.
\item The raw data is binned into spatial pixels substantially 
smaller
than $L/10$, so that this pixelization process retains all 
the information except that which was already captured by 
the traditional method.
\item The linear (KL) method is used to measure anisotropy 
parameters
such as $\Omega^{0.6}/b$ (from redshift-space distortions), 
a residual extinction template and corrections to the radial 
selection function, as well as large-scale band powers. 
\item Uncorrelated estimates of the power spectrum on scales
$k^{-1}\simgt L/10$ are computed with the quadratic method,
extending down to even smaller scales if 
$N\sim 10^4-10^5$ pixels are feasible.  This can be done both by
incorporating the systematic effects found in Step 3 with the KL
method, and by using quadratic estimators which are insensitive to these
systematic effects.  The comparison of the results for the band powers
allows us to quantify how successful we are in eliminating these
effects. 
\item The entire process may be iterated, using the (smoothed) 
measured power
spectrum as the fiducial one.
\item Remaining cosmological parameters are estimated with a 
likelihood 
or $\chi^2$ 
analysis from the power spectrum. 
\end{enumerate}
This approach should allow future redshift surveys to
realize their full potential to constrain cosmological models.
In the meantime, it appears worthwhile to reanalyze various
existing surveys with the same pipeline, to eliminate any 
method-induced artifacts and allow more accurate cross-comparisons 
of results.

\acknowledgements 
We thank the Aspen Center for
Physics where this paper neared its final form,  
David Weinberg and Josh Frieman for useful conversations, and an
anonymous referee for a very close reading of the manuscript and many
helpful suggestions.  
Support for this work was provided by NASA through grant NAG5-6034 
and Hubble Fellowships
HF-01078.01-94A and HF-01084.01-96A, awarded by the Space Telescope
Science Institute, which is operated by AURA, Inc. under NASA contract
NAS5-26555.  M.A.S. acknowledges the support of the Alfred P. Sloan
Foundation, Research Corporation, and NSF Grant AST96-16901.  
 

\appendix

\twocolumn

\setcounter{secnumdepth}{3}

\section{APPENDIX A: SHOT NOISE REMOVAL}


There are two basic approaches in the literature to removing shot noise.
In this Appendix,
we show that the two give essentially identical results. 

\subsection{Using the Gaussian approximation}

\label{GaussNoiseSec}

Let us define write our shot noise corrected band power estimator as
$(q_\i-\bt_\i)$, where $\bt_i$ denotes our  {\it bias correction}.
\Eq{GenQuadExpecEq} shows that this estimator will only be unbiased 
if $\expec{\bt_i}=b_\i$, where $b_i$ is given by \eq{GenQuadBiasEq}.
A convenient way of removing the noise bias in pixelized
methods is to choose simply $\bt=b_i$, since
it can be done after pixelizing, without ever going back 
to the individual galaxy positions;
for a general quadratic combination
of pixels $q_\i$ as in \eq{EmatrixDefEq}, using \eq{xCovEq} shows that
the shot noise bias can be written in terms of 
pixelized quantities alone, as 
\beq{PixelizedNoiseBiasEq}
b_\i=\tr[\E_\i\N],
\eeq
where $\N$ is given by \eq{NdefEq}. 

\subsection{The strict minimum variance method}

In the approximation that the shot noise 
fluctuations in the pixels has a Gaussian probability 
distribution, the above-mentioned method of choosing 
$\bt_i=b_i$ is readily shown to give the unbiased power estimator
with the smallest variance. 
This is an excellent approximation 
when the number of galaxies is large, as we now show. 
The strict minimum-variance method is (Peebles 1980) to simply omit
self-pairs in \eq{DirectQuadEq}: 
\beq{DirectQuadEq2}
q_\i-\bt_\i = 
\sum_{\a\ne\b} {E_\i(\r_\a,\r_\b)\over\nbar(\r_\a)\nbar(\r_\b)}.
\eeq
This corresponds to the shot noise correction
\beq{StrictNoiseBiasEq}
\bt_\i = \int E_\i(\r,\r) {n(\r)\over\nbar(\r)^2}d^3r 
= \sum_\a {E_\i(\r_\a,\r_\a)\over\nbar(\r_\a)^2},
\eeq
which is to be compared with \eq{GenQuadBiasEq}.
This is an unbiased method since $\expec{\bt_i}=b_i$.
How much smaller is the variance with this approach?
We illustrate this with the toy problem of estimating the  
power at $k=0$ in a volume-limited survey with 
$N$ galaxies, which is simply proportional to 
\beq{}
q \equiv (N-\Nbar)^2-\bt,
\eeq
where $N$ is a Poisson-distributed 
random variable with mean $\Nbar$.
Since $\expec{(N-\Nbar)^2}=\Nbar$, the shot noise 
correction $\bt_i=b_i$
corresponds to the choice $\bt=\Nbar$. The strict minimum variance
method of \eq{StrictNoiseBiasEq} corresponds to $\bt=N$.
Both methods are unbiased, giving $\expec{q}=0$.
The higher order moments differ, however.
The variances $\expec{q^2}$ are $2\Nbar^2$ and
$2\Nbar^2+\Nbar$, respectively, so whereas the strictly optimal
method gives the same variance that Gaussian noise would, 
the variance of the other method is a factor $(1+1/2\Nbar)$ larger.
\footnote{
The differences between the two methods 
get larger for higher moments, but always remain 
of the order $1/\Nbar$. Compared with the Gaussian approximation
$\expec{q^3}=8\Nbar^3$, the skewness of the two methods is up 
by factors of $(1+2/\Nbar)$ and $(1+11/4\Nbar+1/8\Nbar^2)$,
and compared with the Gaussian approximation
$\expec{q^4}=60\Nbar^4$, the kurtosis is up 
by factors of $(1+12/5\Nbar+2/15\Nbar^2)$ and 
$(1+33/5\Nbar+23/12\Nbar^2+1/60\Nbar^3)$.
}

\subsection{Which method is preferable?}

This means that the optimal method only reduces the standard deviation
by a negligible $0.01\%$ for $N=10^4$ galaxies. Moreover, it is
incorrect to claim that the strictly optimal method in some
sense removes the ``exact'' shot noise: since the higher order
moments depart from the Gaussian values even for this method, 
there is clearly Poissonian noise left in $q$. 
The difference in variance between the two methods  
remain equally negligible for more realistic examples, generally
being of order the inverse of the number of galaxies in the survey.
The choice of which method to use should therefore be dictated by
practical convenience. Whereas the strict minimum variance method is 
of course trivial 
to implement in techniques involving an explicit sum over galaxy pairs 
(such as the FKP method),
the other method is generally simpler to use 
for pixelized techniques, since it can be implemented using 
the pixelized data alone.\footnote{There is one useful exception: 
one can use the strict minimum variance method based on pixelized data alone 
in the special case where the pixels are counts in (sharp-edged) cells,
in which case the terms $x_i^2$ get replaced by $x_i(x_i-1)$.}

\section{APPENDIX B: DERIVATION OF INTEGRAL CONSTRAINT AND RELATED
EXPRESSIONS}
\label{IntegralConstraintMathSec}
In this appendix, we derive some of the results described in the text:
the integral constraint correction for the traditional Fourier method
(\Sec{TraditionalSec}), and the optimal way to immunize the data from the 
untrustworthy modes of \Sec{SystematicsSec}. 

\subsection{The Integral Constraint in the Direct Fourier Method}

\label{ContinuousIntConSec}

How should one deal with the integral constraint when using a traditional 
method as in \sec{TraditionalSec}? From 
the discussion in \Sec{SystematicsSec}, it is clear that we
should modify the  
weighting functions of \eq{FourierPixelizationEq} so that 
they become orthogonal to the mean density, {\ie}, so that
$\int\psi_i(\r)d^3r=0$.
There are infinitely many ways of doing this, and some are clearly better than 
others if we want the power estimators $q_i=|x_i|^2$ to retain as much cosmological 
information as possible. We here derive one such correction method which
is both simple and intuitive, following Tegmark (1997c).  
We will adopt a more ambitious approach in the following subsection,
deriving the optimal correction method for the pixelized case.
At the end of this Appendix, we show that in the
small-scale limit, the two methods are in fact identical. 

If we use our guess $\nbar_0$ (equation~\ref{etaDefEq}) 
in place of the unknown true selection function $\nbar$ in \eq{xDefEq},
we will have $\expec{x_i}=(\nnorm-1)\psih_i(0)\neq 0$.
When using a traditional power estimator $q_i=|x_i|^2$, this causes 
a systematic positive power bias $(\nnorm-1)^2|\psih_i(0)|^2$ that
we cannot subtract off, as $\nnorm$ is unknown.
We must therefore modify $\psi_i$ so that $\psih_i(0)$ vanishes.
Let $\nnormh$ denote our estimate of $\nnorm$.
We will choose $\nnormh$ so that this bias vanishes, {\ie}, 
so that the integral constraint
\beq{ImplicitEtaDef}
\int\left[{n(\r)\over\nnormh\nbar_0(\r)}-1\right]\phi(\r) 
d^3r=0
\eeq
holds, or explicitly, 
\beq{ExplicitEtaDef}
\nnormh\equiv {1\over\phih(\vzero)} 
\int {n(\r)\over\nbar_0(\r)}\phi(\r) d^3r.
\eeq
This is an unbiased estimator of the density 
normalization, since 
$\expec{\nnormh}=\nnorm$, the true value.
Substituting $\nbar(\r)=\nnormh \nbar_0(\r)$ 
and equations\eqn{FourierPixelizationEq}
and\eqn{ExplicitEtaDef}
into \eq{xDefEq}, we obtain 
\beqa{Fh2DerivEq}
x_i
&=&
{1\over\nnormh}
\left[\int{n(\r)\over\nbar_0(\r)}e^{i\vk_i\cdot\r}\phi(\r) 
d^3 r
-\phih(\k_i)\nnormh\right]\nonumber\\
&=&
{1\over\nnormh}
\left[\int{n(\r)\over\nbar_0(\r)}e^{i\vk_i\cdot\r}\phi(\r) 
d^3 r
-
{\phih(\k_i)\over\phih(\vzero)}\int{n(\r)\over\nbar_0(\r)}
\phi(\r) d^3 r\right]
\nonumber\\
&=&
{\nnorm\over\nnormh}\int {n(\r)\over\nbar(\r)}\psi_i(\r) d^3 
r
\approx\int {n(\r)\over\nbar(\r)}\psi_i(\r) d^3 r,
\eeqa
where the function $\psi_i$ is defined by
\beq{psiiDefEq}
\psi_i(\r) \equiv
\left[e^{i\vk_i\cdot\r}-
{\phih(\k_i)\over\phih(\vzero)}\right]\phi(\r).
\eeq
Hence its Fourier transform is 
\beq{psiihatEq}
\psih_i(\k) = 
\phih(\k-\k_i) - {\phih(\k_i)\over\phih(\vzero)}\phih(\k).
\eeq
The relative error in $\nnormh$ is of order the inverse square root of
the number of galaxies in the survey, so 
we can to a good approximation
treat $\nnorm$ as a known constant from here on and take
$\nnorm/\nnormh=1$ on the last line of \eq{Fh2DerivEq}.
We see that the volume weighting $\psi_i$ given by 
\eq{psiiDefEq} is better than the traditional choice 
of \eq{FourierPixelizationEq} since it is orthogonal to the
mean, \ie, it satisfies $\psih_i(\vzero)=0$, which guarantees 
that $\expec{x_i}=0$. 

In practice, we need never use \eq{psiiDefEq} to
compute $x_i$ with \eq{xDefEq}, since this is
implicitly done if
we first correct $\nbar$ by estimating its normalization with 
\eq{ExplicitEtaDef} and then apply the simple weight function of 
\eq{FourierPixelizationEq}. (This is mathematically equivalent to
applying the weight function \eq{psiiDefEq}
directly to the data, using an arbitrary 
$\nbar$-normalization.)
However, we do need \eq{psiiDefEq} to
derive the expressions for the
shot noise correction and normalization given in 
Equations\eqn{sshotDefEq} and\eqn{NormDefEq}.
Substituting \eq{psiiDefEq} into \eq{GenQuadBiasEq}
gives the shot noise correction 
\beqa{shotCorrEq1}
b_i&=&\int{|\psi_i(\r)|^2\over \nbar(\r)} d^3 
r\nonumber\\
&=&
\int\left|e^{i\vk_i\cdot\r}-
{\phih(\k_i)\over\phih(\vzero)}\right|^2
{\phi(\r)^2\over\nbar(\r)}d^3r,
\eeqa
and expanding the square completes our derivation of 
\eq{sshotDefEq}.
The normalization coefficient $A_i$ of 
\eq{PtDefEq} is determined by the requirement that the
window function integrate to unity, {\ie},
$A_i = \int|\psih_i(\k)|^2 d^3 k/(2\pi)^3$.
Using Parseval's theorem, we obtain
\beqa{NderivationEq2}
A_i&=&\int |\psi_i(\r)|^2 d^3 r\nonumber\\
&=&\int\left|e^{i\vk_i\cdot\r}-
{\phih(\k_i)\over\phih(\vzero)}\right|^2
\phi(\r)^2 d^3r,
\eeqa
and expanding the square as above completes our derivation of 
\eq{NormDefEq}.

\subsection{How important is this correction?}

Let us evaluate the integral constraint correction factor 
$A_i$ 
for a couple of simple examples. We first note that for the 
special case of \eq{psiChoiceEq}, we 
have $\phi(\r)^2\propto \phi(\r)$. Hence $a(\k)\propto
\phih(\k)$,
and \eq{NormDefEq} reduces to 
\beq{ParkApproxEq}
A_i=\left(1-
\left|{\phih(\vk_i)\over\phih(\vzero)}\right|^2\right)a(\vzero),
\eeq
which we recognize as the result of 
Park {\etal} (1994).
For volume-limited surveys, 
the prescriptions given by equations\eqn{psiChoiceEq}, \eqn{APMeq} and\eqn{FKPeq} 
all coincide, so this 
expression is exact for the volume-limited case with these galaxy 
weighting schemes.
For flux-limited surveys, on the other hand, 
these schemes all give a decreasing weight function $\psi$,
since $\nbar$ decreases with distance.
For the simple Gaussian case 
$\phi(\r)=\exp[-(r/R)^2/2]/\pi^{1/4}R^{1/2}$,
\eq{NormDefEq} gives
\beq{GaussianNeq}
A_i = 1 + e^{-(Rk_i)^2} - 2e^{-{3\over 4}(Rk_i)^2},
\eeq
whereas the approximation\eqn{ParkApproxEq} gives 
\beq{ParkNeq}
A_i = 1 - e^{-(Rk_i)^2}.
\eeq
A Taylor expansion shows that for $kR\ll 1$, 
the latter overestimates $A_i$ by a factor of two
as illustrated in Figure 2.

\bigskip

\noindent
\centerline{\rotate[r]{\vbox{\epsfxsize=6.5cm\epsfbox{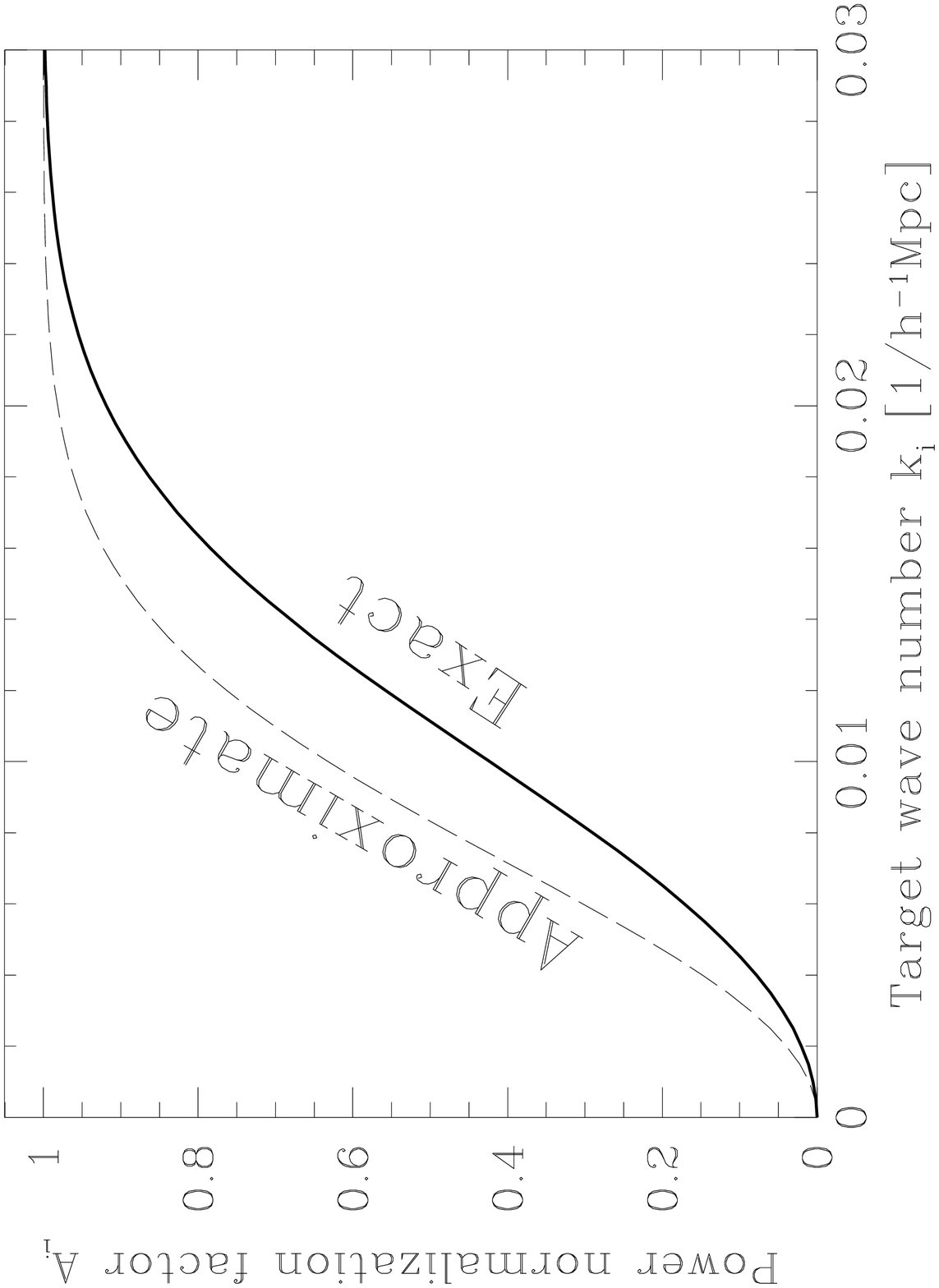}}}}
{\footnotesize {\bf FIG. 2}
--- The exact expression for the integral constraint correction $A_i$ 
is plotted together with the
approximation of Park {\it et al} (1994) for a Gaussian 
weight function $\psi(\r)\propto\exp[-(r/R)^2/2]$,
$R=100 h^{-1}$Mpc.
}

\clearpage
\subsection{Eliminating contaminated modes with pixelized methods: A
Simple Solution} 

In this and the next subsection, we continue the discussion of
\Sec{SystematicsSec}, and present a method to project out modes of
the density field that we believe might be contaminated by, \eg,
errors in the selection function or in the assumed extinction map.
Our starting point is \eq{ZdefEq}, the $N\times M$ matrix $\Z$
of untrustworthy modes. 

To remedy the problem, we construct a new ``cleaned" data set 
that is independent of $\vnnorm$, the coefficients of these modes. Let
us define  
\beq{xDefEq2}
\x\equiv\PP\x',
\eeq
where $\PP$ is an $N\times N$ matrix satisfying
\beq{PPorthoEq}
\PP\Z=\vzero, 
\eeq
{\ie}, having the columns of $\Z$ in its null space. 
This implies that $\PP$ has at most rank $N-M$.  We will choose it to
have exactly this rank, since otherwise $\PP$ will destroy more 
information than necessary 
(for instance, null matrix choice $\PP=\vzero$ satisfies 
\eq{PPorthoEq}, but destroys all our information).
It is easy to construct such matrices, a simple choice being 
\beq{SimplePPeq}
\PP=\I - \Z(\Z\dag\Z)^{-1}\Z\dag.
\eeq
This is the Hermitean ($\PP\dag=\PP$) projection matrix 
($\PP^2=\PP$) projecting onto the subspace orthogonal
to the columns of $\Z$.
Our corrected data set $\x$ satisfies 
\eq{xExpecEq}, since
$\expec{\x}=\PP\Z\vnnorm =\vzero$.
Letting $\C'$ denote the covariance matrix of the uncorrected
data set, the corrected data will have the covariance matrix
\beq{xCovarEq2}
\C\equiv\expec{\x\x^\dagger}=\PP\C'\PP\dag.
\eeq
Once $\x$ and $\C$ have been computed, the rest of the 
pixelized 
analysis proceeds just as described in Sections~\ref{BruteSec},
\ref{KLsec} or~\ref{QuadSec}.
The only complication is that $\C$ is now singular, 
having rank $N-M$ instead of $N$.
As shown in the Appendix of T97, the correct way to deal with
this in the quadratic method is to replace all occurrences of $\C^{-1}$
(which is of course undefined) by the ``pseudo-inverse"
of $\C$, defined as 
\beq{PseudoInverseEq}
\PP\left[\C+\gamma\Z\Z\dag\right]^{-1}\PP
\eeq
for some constant $\gamma\neq 0$.  T97 shows that the result is independent 
of $\gamma$, and that a good choice for numerical stability
is $\gamma\sim c/N$, where $c$ is the order of magnitude
of a typical matrix element of $\C$. The same trick can be used for the
KL method, in the step where \eq{EigenEq2} is reduced to an ordinary eigenvalue problem
by Cholesky decomposing $\C$ as described in TTH.

\subsection{The optimal solution}

\Eq{SimplePPeq} does not give the only rank $N-M$ projection 
matrix satisfying $\PP\Z=\vzero$ --- there are in fact infinitely many
such matrices of the form $\PP=\I-\Z(\Z\dag\M\Z)^{-1}\Z\dag\M$, 
where $\M$ is an arbitrary non-singular matrix, and \eq{SimplePPeq}
simply corresponds to the case $\M\propto\I$.
Since they all have the same null space $\Z$, it is clear that they all
destroy the same information (all the information 
about the untrustworthy modes, no more and no less). 
The power spectrum Fisher matrix for the cleaned data set is therefore
independent of $\M$, so our choice is purely one of numerical convenience.
For the quadratic method in particular, there turns out to be a much
more appropriate choice than that of \eq{SimplePPeq}, which altogether eliminates
the above-mentioned problem of $\C$ being singular by allowing the 
quadratic pair weighting $\E$ to be computed analytically.
We will derive this choice of $\PP$ by generalizing the derivation of the
quadratic method (T97) to include our constraint that the results be
independent of the corrupted modes.

The most general quadratic estimator can clearly be written
as in \eq{EmatrixDefEq}
for some symmetric matrix $\E_i$.
As shown in T97, this implies that the variance of $q_i$ is given by
\beq{qiVarEq}
V(q_i)\equiv\expec{q_i^2}-\expec{q_i}^2={1\over 2}\tr[\C\E_i\C\E_i]
\eeq
and that the signal, the expected contribution to $q_i$ from the power 
band of interest, is
$\tr[\C,_i\E_i]/2$,
where $\C,_i$ is defined by \eq{CpDefEq}. To 
maximize the signal-to-noise ratio, we want to 
minimize the variance given a fixed signal, {\ie}, subject to the
constraint that $\tr[\C,_i\E_i]$ is held constant.
If we write
\beq{qexpandEq}
q_i = {1\over 2} (\x + \Z\vnnorm)^\dagger\E_i(\x + \Z\vnnorm),
\eeq
it is clear that we can phrase our constraint on $\E_i$ as $\E_i\Z=\vzero$.
This is in fact 
$N\times M$ separate constraint equations, so using the fact that 
$\E_i$ is symmetric, 
our constrained minimization problem involves minimizing
\beq{LagrangeEq}
L \equiv \tr\left\{{1\over 2}\C\E_i\C\E_i-\lambda\C,_i\E_i+\lambda[\Z\A\dag+\A\Z\dag]\E_i\right\},
\eeq
where $\A$ is some arbitrary $N\times M$ matrix of Lagrange multipliers.
Requiring the derivatives with respect to the components of $\E_i$ to vanish, 
we obtain
\beq{LagrangeSolEq}
\E_i\propto\C^{-1}\left[\C,_i - \Z\A\dag - \A\Z\dag\right]\C^{-1},
\eeq
where $\A$ is determined by the constraint $\E_\i\Z=\vzero$. 
Defining $\Zt\equiv\C^{-1}\Z$, the solution is 
\beq{Aeq}
\A = \left[\I-{1\over 2}\Z(\Zt\dag\Z)^{-1}\Zt\dag\right]\C,_i\Zt(\Zt\dag\Z)^{-1},
\eeq
which can be verified by direct substitution.
Substituting this back into \eq{LagrangeSolEq} finally yields
\beq{GenQuadMethodEq}
\E_i\propto\PP\dag\C^{-1}\C,_i\C^{-1}\PP,
\eeq
where 
\beq{OptPPeq}
\PP=\I - \Z(\Z\dag\Zt)^{-1}\Zt\dag = \I - \Z(\Z\dag\C^{-1}\Z)^{-1}\Z\dag\C^{-1}
\eeq
is a projection matrix satisfying $\PP\Z=\vzero$, $\PP\dag\Zt=\vzero$ and
$\C^{-1}\PP=\PP\dag\C^{-1}$.
Inserting \eq{GenQuadMethodEq} into \eq{EmatrixDefEq}, we
see that our quadratic estimator retains 
the simple form of \eq{yDefEq2} if we generalize \eq{zDefEq} to 
\beq{zDefEq2}
\z\equiv\C^{-1}\PP\x,
\eeq
so after cleaning the data set (replacing $\x$ by $\PP\x$), 
the quadratic method proceeds exactly as before.
This generalization of the quadratic method clearly reduces to the prescription
in \sec{QuadSec} ($\z=\C^{-1}\x$) if
there are no untrustworthy modes, in which case $M=0$ and $\PP=\I$.
Note that this technique is quite useful for estimating the power
spectrum from cosmic microwave background experiments as well, in which case
obvious candidates for corrupted modes are the monopole and 
the three dipole components.

\subsection{Relation between the pixelized and continuous cleaning schemes}

In this section, we will show that the integral correction procedure for
traditional methods that we derived above is lossless for the FKP 
volume weighting in the small-scale limit, corresponding to the quadratic method.

When we have merely one untrustworthy mode ($M=1$) corresponding to the 
normalization of $\nbar$, the matrix defined by
\eq{ZdefEq} consists of a single column vector; $\Z=\u$.
Using the continuum pixels $x(\r)$ of \eq{ContinuumPixelEq}, this vector is simply
the $\k=\vzero$ (constant) mode, {\ie}, 
\beq{uDefEq}
u(\r)=1.
\eeq
Let us now evaluate the optimal power estimate $q_i$ that we derived above, given 
by Equations~\eqn{yDefEq2} and\eqn{zDefEq2}.
Since $\U=\u$, we have
\beq{PxEq}
\PP\x = \x - \left({\u\dag\C^{-1}\x\over\u\dag\C^{-1}\u}\right)\u,
\eeq
where Equations\eqn{ContinuumCinvEq} and\eqn{uDefEq} show that 
$\u\dag\C^{-1}\x=\int\phi(\r)x(\r)d^3r$ and $\u\dag\C^{-1}\u=\int\phi(\r)d^3r=\phih(0)$.
We can thus write
\beq{CPxEq2}
z(\r) = (\C^{-1}\PP\x)(\r) = 
\int\left[\dirac(\r-\r')-{\phi(\r)\over\phih(0)}\right]\phi(\r')x(\r')d^3r'.
\eeq
Fourier  transforming this with respect to $\r'$, 
\Eq{yDefEq3} now shows that $q_i=|x_i|^2/2$, where
\beq{zhEq}
x_i = \zh(\k_i) = \int\psi_i(\r)x(\r) d^3r,
\eeq
and the volume weighting function $\psi_i$ is exactly the one that we derived
in \sec{ContinuousIntConSec}, given by \eq{psiiDefEq}. In other
words, we have shown that the FKP choice of the function $\phi$
together with the volume weighting of \eq{psiiDefEq} is identical
to the lossless quadratic method in the small-scale limit.

\section{APPENDIX C: REDSHIFT DISTORTIONS AND CLUSTERING EVOLUTION}

In \sec{SystematicsSec} and Appendix B, we showed that the pixelized
methods allow a  
more 
ambitious approach than is feasible with the direct Fourier 
methods,
incorporating multiple integral constraints, since all the 
complications simply became buried in the appropriate 
matrices. 
In this Appendix, we show how two additional complications 
can be incorporated
with pixelized methods in 
the same vein: redshift distortions and clustering evolution.

\subsection{Redshift distortions}

A ubiquitous problem with power spectrum estimation is that 
of 
``redshift distortions''. 
When estimating the  distance to a galaxy by its 
redshift,
galaxies receding faster than the Hubble flow due to local 
gravitational interactions appear to be further away than 
they
really are, and vice versa. This was first discussed by 
Kaiser (1987) in the context of $P(k)$,
and a recent review is given by Hamilton (1997c).
Denoting the apparent density field in redshift space
$\delta_s(\r)$, Hamilton \& Culhane (1996) use Kaiser's formalism to show 
in linear perturbation theory that 
\beq{RedshiftSpaceEq1}
\delta_s = 
\left[1 + \beta \left({\partial^2\over\partial r^2}
      + {\alpha(\r)\over r} {\partial\over\partial r}\right)\nabla^{-2}\right]\delta_r,
\eeq
where $\beta\equiv \Omega^{0.6}/b$, the constant 
$b$ is the so-called linear bias factor, and
\beq{alphaDefEq}
\alpha(\r)\equiv 2+{\partial\ln\nbar(\r)\over\partial \ln r}
\eeq
is two plus the logarithmic slope of the radial selection
function. 
Fourier transforming this gives (Hamilton \& Culhane 1996;
Hamilton 1997c)
\beq{zspaceEq}
\deltash(\vk) = \delth(\vk) + 
\beta\int f(\vk,\vk')
\delth(\vk') {d^3k'\over(2\pi)^3},
\eeq
where the function $f$ is defined by
\beq{fDefEq}
f(\vk,\vk')\equiv
\int e^{i(\vk'-\vk)\cdot\r}
\left[(\vkh'\cdot\rh)^2 - {\alpha(\r)i\over 
k'r}(\vkh'\cdot\rh)\right]d^3r.
\eeq
Thus we obtain 
\beq{zspaceCorrEq}
\expec{\deltash(\vk)^*\deltash(\vk')}
= (2\pi)^3\int g(\vk,\vk',\vk'') P(\vk'') d^3k'',
\eeq
where 
\beqa{zspaceCorrEq2}
&&g(\vk,\vk',\vk'')
\equiv\dirac(\vk-\vk'')\dirac(\vk'-\vk'')+\nonumber\\
&+&\beta[\dirac(\vk-\vk'')f(\vk',\vk)+\dirac(\vk'-\vk'')f(\vk,\vk')^*]+\nonumber\\
&+&\beta^2 f(\vk,\vk'')^* f(\vk',\vk'').
\eeqa
The above expressions are derived and discussed in detail
by Zaroubi \& Hoffman (1996), and also in Tegmark \& Bromley (1995) and
T95 for the volume limited case; see Szalay, Matsubara, \& Landy
(1997) for further discussion. 
The key point here is that although 
$\expec{\deltash(\vk)^*\deltash(\vk')}$
is no longer diagonal, and 
rather messy,
it is still linear in the power spectrum. 
Thus the pixel covariance matrix $\C$
will still be some shot noise term plus a term {\it linear} 
in $P(\vk)$.
In other words, by letting $\C_{,i}$ refer to the derivative 
of
$\C$ with respect to the band powers in {\it real} space
instead of redshift space, the quadratic method will measure 
the real 
space power spectrum directly (given {\it a priori}
knowledge of $\beta$), and the corresponding window 
functions (the rows of $\F$, say) will show the contributions 
to the
measurements $q_i$ from the various real space power bands.

\subsection{Clustering evolution}

The density fluctuation field $\delt$
maintains its shape in linear perturbation theory, simply 
increasing in amplitude by a position-independent growth 
factor $D$.
Since we are seeing distant galaxies at an earlier time,
we see the apparent density fluctuations 
\beq{EvolutionEq}
\delta_a(\r) \equiv D(r) \delt(\r),
\eeq
where $D(r) = 1/(1+z)$ for $\Omega=1$.
This effect is straightforward to 
include in
a pixelized analysis.
\Eq{NdefEq} remains unchanged and \eq{SdefEq} simply gets 
replaced by
\beq{SdefEq2}
\S_{ij}=\int \psih'_i(\vk)\psih'_j(\vk)^*
P(k)\dV,
\eeq
where we have defined the functions  
$\psi'_i(\r)\equiv D(r)\psi_i(\r)$.
This correction is quite small for shallow 
galaxy surveys, where $\nbar$ typically varies 
dramatically between $z=0$ and $z=0.2$, a range over which
$D$ changes by at most about $20\%$, less for small $\Omega$.  If this
effect is incorporated into the analysis, $\Omega$ can be made a free
parameter to be fit for in the pipeline. 

This clustering evolution should not be confused with
galaxy evolution, which we do not discuss here, and which affects only
$\nbar$, not $\delt$.

\section{APPENDIX D: NORMALIZATION CONVENTIONS}

Unfortunately, the power spectrum $P(k)$ is defined
in many different ways in the literature, differing by 
normalization factors such as $(2\pi)^3$ and a fiducial 
box volume $V$. 
In this paper, we normalize Fourier transforms as
\beq{FourierNormEq}
\hat{f}(\k)\equiv \int f(\r) e^{-i\vk\cdot\r} d^3 r,
\eeq
and normalize $P(k)$ so that 
\beq{PnormEq}
\expec{\delth(\k)^*\delth(\k')} = (2\pi)^3 \dirac(\k-\k') 
P(k).
\eeq
The units of $P(k)$ are volume.
With this normalization, the dimensionless power
$\Delta^2$ of Peacock \& Dodds (1994) is given by
\beq{DeltaEq}
\Delta^2(k) = {4\pi\over(2\pi)^3} k^3 P(k),
\eeq
the {\rms} fluctuations $\sigma_8$ in a sphere of radius
$R=8 h^{-1}$ Mpc are
\beq{sigma8eq}
\sigma_8^2 ={4\pi} \int_0^\infty 
\left[{\sin x-x\cos x\over x^3/3}\right]^2 P(k) {k^2 dk\over(2\pi)^3},
\eeq
where $x\equiv kR$, and the
Sachs-Wolfe quadrupole $Q$ in the cosmic microwave background
is given by
\beq{SWeq}
Q^2\equiv {5\over 4\pi}C_2 = {10\over\pi^2}\int_0^\infty 
{j_2(x)^2\over x^4} P(k) k^2 dk,
\eeq
where $x\approx 2kc/H_0\approx k\times 6000 h^{-1}$ Mpc
and 
\beq{j2eq}
j_2(x) = {3\sin x - 3 x \cos x - x^2 \sin x\over x^3}.
\eeq
T95 uses a convention where the $(2\pi)^3$ factor in 
\eq{PnormEq} is replaced by $(2\pi)^6$.

\clearpage
\onecolumn
\begin{figure}[phbt]
\centerline{{\vbox{\epsfysize=20cm\epsfbox{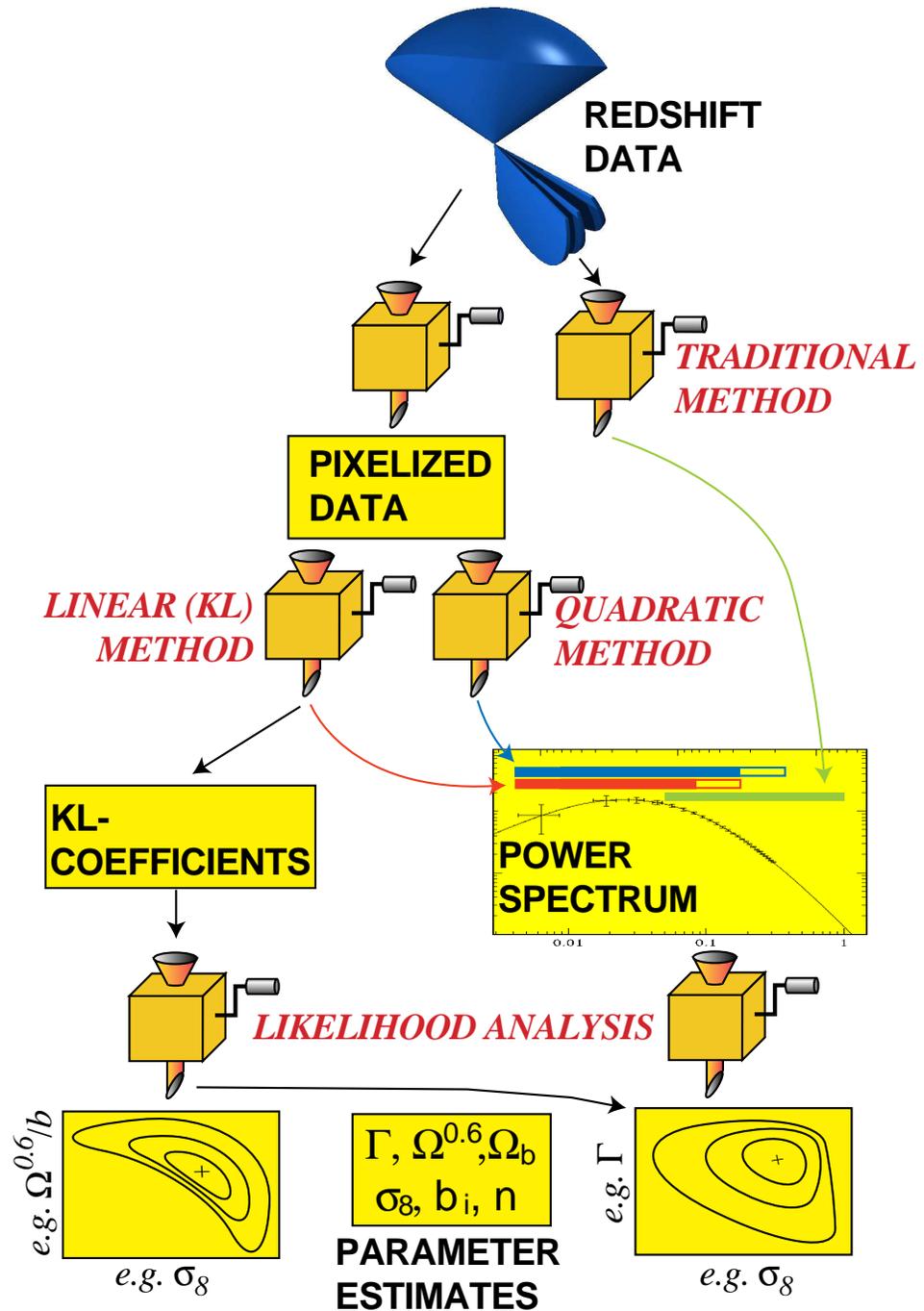}}}}
\smallskip
\caption{
\label{PipelineFig}
We propose analyzing large future galaxy redshift surveys 
such as 
the SDSS, using three techniques in conjunction:
a traditional Fourier approach
on small scales, a pixelized quadratic matrix method on large
scales and a pixelized Karhunen-Lo\`eve eigenmode
analysis to probe anisotropic effects such as
redshift-space distortions and residual extinction.
The horizontal bars in the power spectrum box indicate that 
the quadratic 
method has a larger dynamic range than the KL method.
The bottom of the figure indicates that numbers 
such as the redshift distortion parameter $\Omega^{0.6}/b$
which reflect anisotropic clustering
can only be optimally constrained using the KL modes, which
retain not merely the overall power, but the phase 
information as well.
}
\end{figure}

\ed
\clearpage
\begin{figure}[phbt]
\centerline{\rotate[r]{\vbox{\epsfysize=15cm\epsfbox{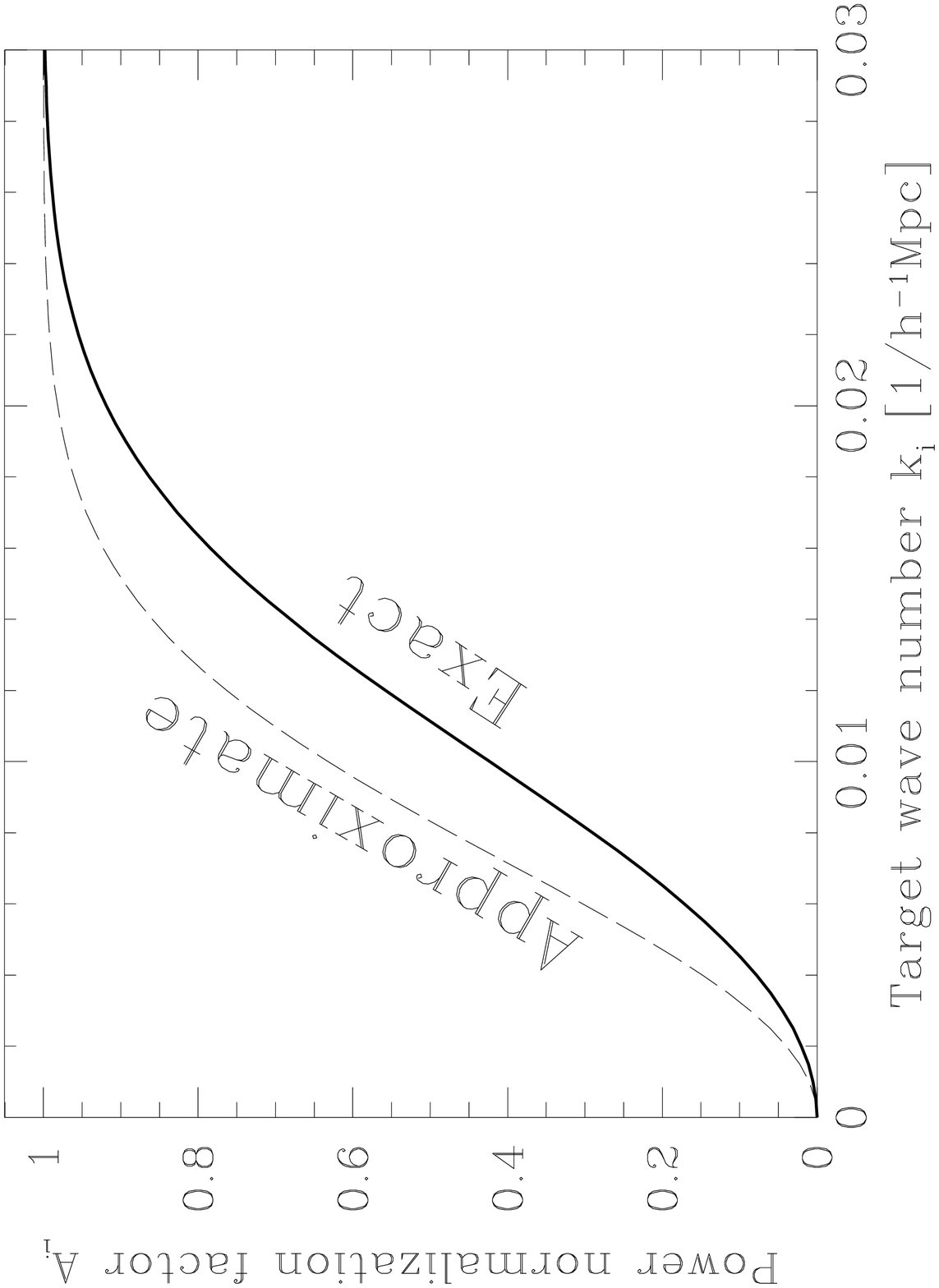}}}}
\smallskip
\caption{
\label{BlowupFig}
The exact expression for the integral constraint correction $A_i$ 
is plotted together with the
approximation of Park {\it et al} (1994) for a Gaussian 
weight function $\psi(\r)\propto\exp[-(r/R)^2/2]$,
$R=100 h^{-1}$Mpc.
}
\end{figure}

\ed